\documentclass[a4paper,11pt]{article}
\pdfoutput=1 

\usepackage{jinstpub} 

\usepackage{here}
\usepackage{threeparttable}
\usepackage{siunitx}
\title{\boldmath ``INTPIX4NA"---new integration-type silicon-on-insulator pixel detector for imaging application}

\author[a,1]{R. NISHIMURA\note{Corresponding author.}}
\author[a]{, S. KISHIMOTO}
\author[b]{, T. SASAKI}
\author[b]{, S. MITSUI}
\author[c]{, M. SHINYA}
\author[a]{, Y. ARAI}
\author[a]{and T. MIYOSHI}

\affiliation[a]{High Energy Accelerator Research Organization (KEK),\\ Oho 1-1, Tsukuba, Ibaraki, 305-0801, Japan}
\affiliation[b]{Kanazawa University,\\ Kakumamachi, Kanazawa, Ishikawa 920-1192, Japan}
\affiliation[c]{Industrial Research Institute of Ishikawa,\\ Kuratsuki 2-1, Kanazawa, Ishikawa 920-8203, Japan}

\emailAdd{ryunishi@post.kek.jp}

\abstract{
INTPIX4NA is an integration-type silicon-on-insulator pixel detector. 
This detector has a 14.1 \si{\times} 8.7 \si{mm^2} sensitive area, 
425,984 (832 column \si{\times} 512 row matrix) pixels and the pixel size is 17 \si{\times} 17 \si{\micro\meter^2}. 
This detector was developed for residual stress measurement using X-rays (the $cos \alpha$ method). 
The performance of INTPIX4NA was tested with the synchrotron beamlines of the Photon Factory (KEK), and the following results were obtained. 
The modulation transfer function, the index of the spatial resolution, was more than 50\% at the Nyquist frequency (\SI{29.4}{cycle/mm}). 
The energy resolution analyzed from the collected charge counts is 35.3\%--46.2\% at \SI{5.415}{k\electronvolt}, 21.7\%--35.6\% at \SI{8}{k\electronvolt}, and 15.7\%--19.4\% at \SI{12}{k\electronvolt}. 
The X-ray signal can be separated from the noise even at a low energy of \SI{5.415}{k\electronvolt} at room temperature (approximately 25--27 \si{\degreeCelsius}). 
The maximum frame rate at which the signal quality can be maintained is 153 fps in the current measurement system. 
These results satisfy the required performance in the air and at room temperature (approximately 25--27 \si{\degreeCelsius}) condition that is assumed for the environment of the residual stress measurement. 
}


\collaboration[c]{on behalf of SOIPIX collaboration}

\begin{document}
\maketitle
\flushbottom

\section{Introduction}
\label{sec:intro}
\subsection{Overview of SOIPIX structure}

\begin{figure}[H]
\centering
\includegraphics[width=\linewidth]{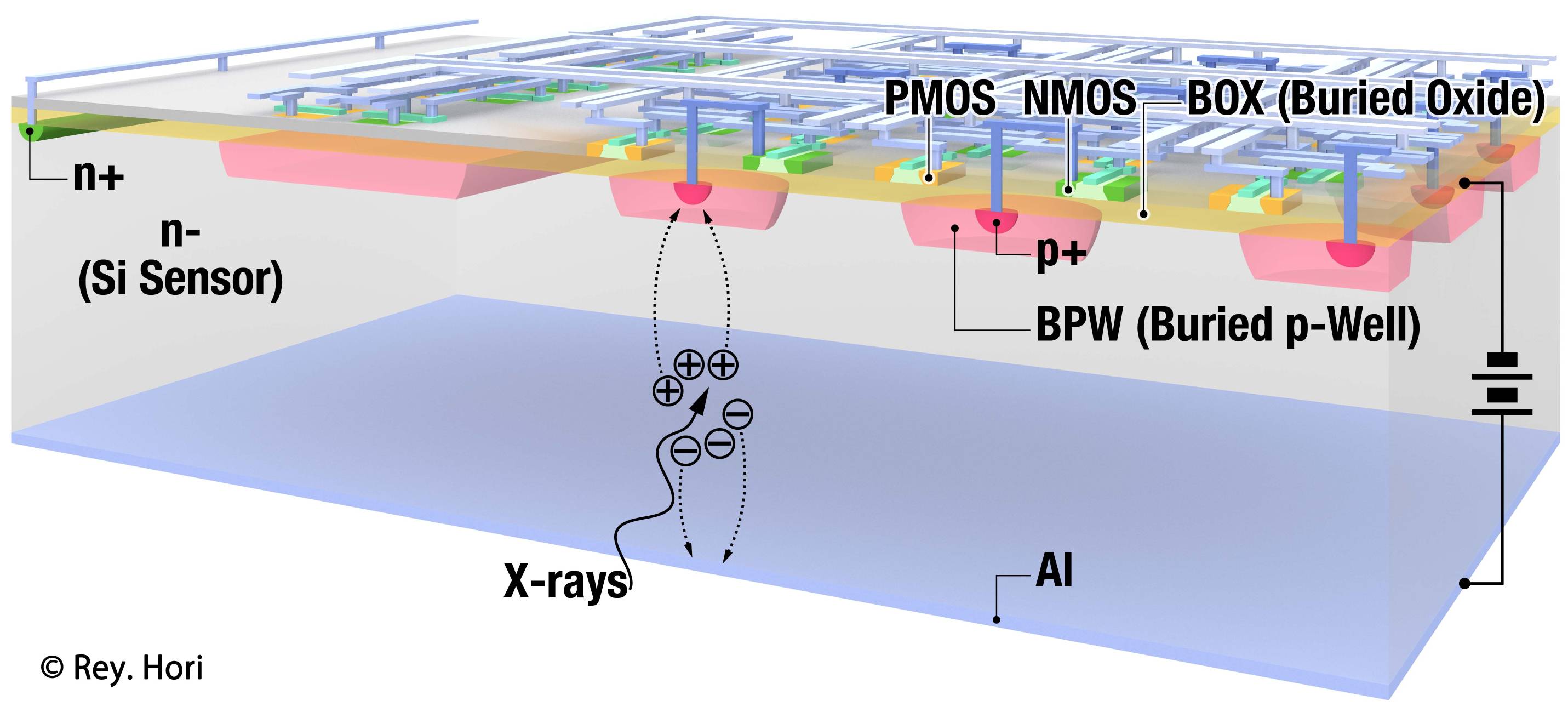}
\caption{Structure of SOIPIX detector.}
\label{fig:soistruct}
\end{figure}

The silicon-on-insulator pixel (SOIPIX) group, led by the High Energy Accelerator Research Organization (KEK), has been developing the SOIPIX, a monolithic pixel detector series using silicon-on-insulator technology. 
The SOIPIX detector is based on a \SI{0.2}{\micro\meter} complementary metal-oxide-semiconductor (CMOS) fully depleted silicon on insulator (FD-SOI) pixel process developed by Lapis Semiconductor Co., Ltd \cite{soi}. 
The SOIPIX detector consists of a thick, high-resistivity Si substrate for the sensor and a thin Si layer for CMOS circuits (figure \ref{fig:soistruct}). 
In the FD-SOI process, transistors were separated from each other and also separated from the Si substrate by an insulator. 
Thus the transistors have lower parasitic capacitance. 
Furthermore, there is no well structure to separate transistors, 
allowing the circuit density to be increased from the bulk CMOS process. 
These characteristics demonstrate that the circuit can be shrunk and operated rapidly at lower power. 
Additionally, the SOIPIX detector has no bump bonding (typically larger than \SI{25}{\micro\meter} bump pitch is required \cite{detrev}) to connect the CMOS circuit and the sensor layer. 
From the above, the SOIPIX detector has the advantage of making a small pixel (the minimum size already developed is 8 \si{\times} 8 \si{\micro\meter^2} \cite{soi5}) with in-pixel signal processing circuits. 

Compared with SOIPIX and other detectors, SOIPIX has a smaller pixel size in direct conversion-type detectors. 
The effective spatial resolution is expected to be higher than that of the indirect conversion-type detector and is expected to reach the upper limit constrained by pixel size. 
A comparison of specifications is shown in table \ref{tab:compare_detectors}. 

\begin{threeparttable}[H]
 \centering
 \begin{tabular}{|c|c|c|c|c|}
 \hline
  & Pixel size & \shortstack{\strut Pixel matrix \\ (1 chip)} & \shortstack{\strut Sensitive area \\ (1 chip)} & \shortstack{\strut Photon detection \\ (conversion type)} \\ \hline
 INTPIX4NA\tnote{*} & 17 \si{\times} 17 \si{\micro\meter^2} & 832 \si{\times} 512 & 14.1 \si{\times} 8.7 \si{mm^2} & Direct (Si) \\ \hline
 \shortstack{\strut X-ray sCMOS \\ (C12849-111U \cite{xscmos1})} & 6.5 \si{\times} 6.5 \si{\micro\meter^2} & 2048 \si{\times} 2048 & 13.3 \si{\times} 13.3 \si{mm^2} & \shortstack{\strut Indirect \\ (Gadolinium \\ oxysulfide (P43) \\ phosphor \SI{10}{\micro\meter})} \\ \hline
 \shortstack{\strut X-ray CCD \\ (S10810-11 \cite{xsccd1})} & 20 \si{\times} 20 \si{\micro\meter^2} & 1500 \si{\times} 1000 & 30 \si{\times} 20 \si{mm^2} & \shortstack{\strut Indirect \\ (CsI deposited \\ on a fiber optic plate \\ (FOP))} \\ \hline
 Eiger 2 R 500K \cite{eiger2} & 75 \si{\times} 75 \si{\micro\meter^2} & 1028 \si{\times} 512 & 77.1 \si{\times} 38.4 \si{mm^2} & Direct (Si / CdTe) \\ \hline
 Medipix 3 \cite{medipix3-1, medipix3-2} & 55 \si{\times} 55 \si{\micro\meter^2} & 256 \si{\times} 256 & 14.1 \si{\times} 14.1 \si{mm^2} & Direct (Si / CdTe) \\ \hline
 \end{tabular}
 \begin{tablenotes}
  \item[*] Detailed specification is shown in table \ref{tab:detector_design}.
 \end{tablenotes}
 \caption{Comparison of X-ray detectors}
 \label{tab:compare_detectors}
\end{threeparttable}

As a part of the application of the SOIPIX detectors, the integration-type SOIPIX detector INTPIX4 \cite{soi3, soi2, soi4} is used for the residual stress measurement of metal surfaces using diffraction X-rays \cite{cosa, cosa2, cosa6}. 
And it was confirmed that this detector can realize a device with superior characteristics, 
compared to the conventional device using an imaging plate, such as improved accuracy, smaller setup size and mesurement. 
However, the INTPIX4 has some issues when applied to the residual stress measurement system, such as the buffer circuit improvement for long-distance transmission, the circuit improvement for higher frame rate, and the reduction of the sensor node reset time. 
Therefore, we decided to develop a new detector, named INTPIX4NA (INTegration-type PIXel detector generation 4 of New Age). 
which is an improved version of the INTPIX4 and is designed for a new X-ray imaging device for residual-stress measurement devices. 

\subsection{Residual stress measurement using $cos \alpha$ method}

\begin{figure}[H]
\centering
\includegraphics[width=\linewidth]{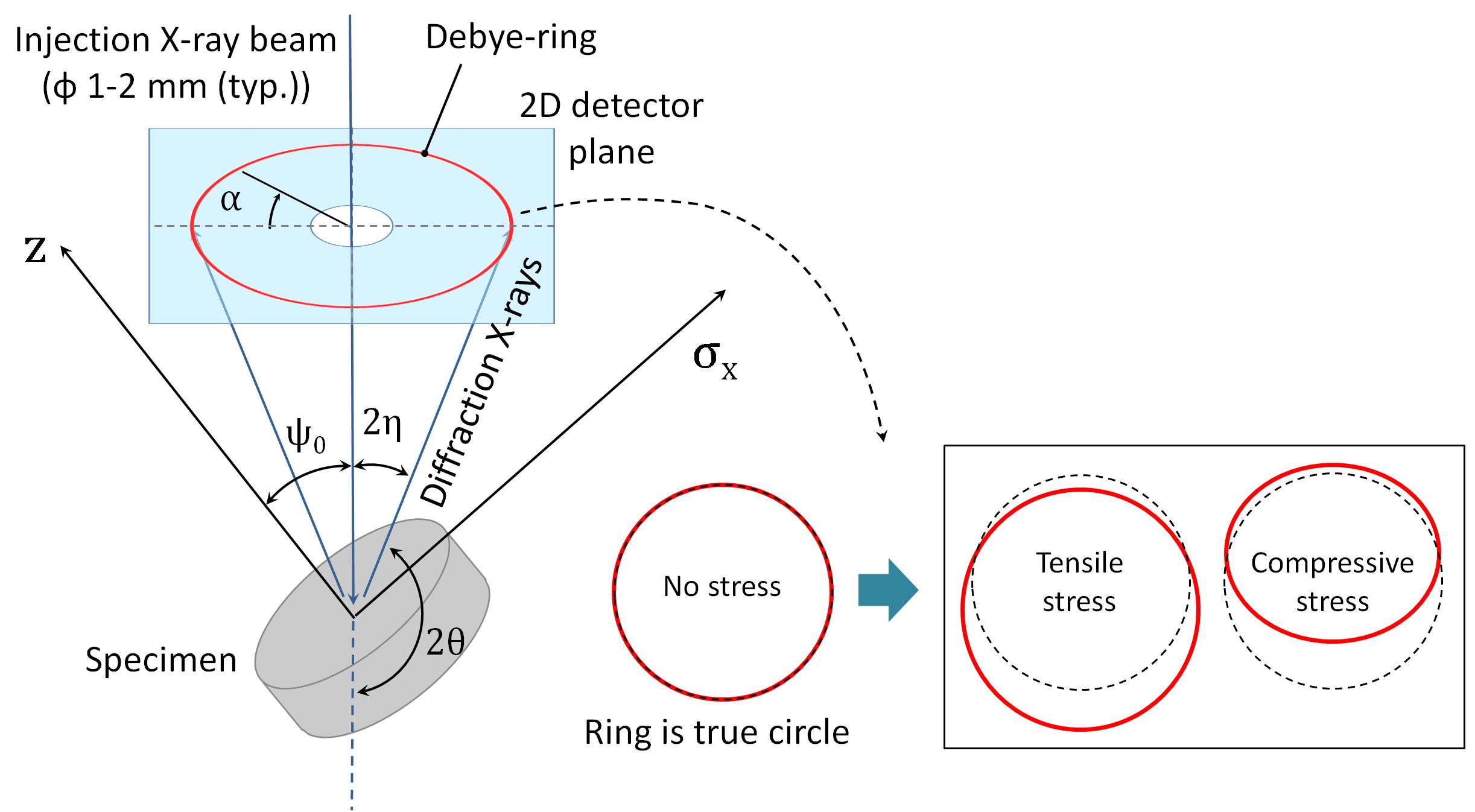}
\caption{Schematic of the relation between the residual stress and the Debye ring deformation.}
\label{fig:cosa}
\end{figure}

The distribution of residual stress is one of the most important factors to evaluate and controls the quality of metal materials in industrial products. 
X-ray diffraction measurements provide one of the most effective methods for evaluating the residual stress without destruction. 
In such measurements, the diffraction of X-ray beams on the surface of a polycrystalline metal forms a Debye ring. 
The effects of residual stress on the crystal structure can be observed through Debye ring deformation (figure \ref{fig:cosa}). 
The $cos \alpha$ method \cite{cosa3, cosa4, cosa5} is a measurement method for analyzing the residual stress from the one-angle image of the Debye ring form. 
In this method, the deformation at each part of the Debye ring is measured, and the relationship between the amount of deformation and the angle of the lattice plane with respect to the stress components is analyzed. 
Thus, the entire shape of the Debye ring (typically more than 2 / 3) is required, and a two-dimensional (2D) X-ray detector is used to measure this. 
The required specification for the detector depends on the scale of the setup and the required stress accuracy. 

The spatial resolution and gain linearity (the linearity of the output response for the collected charge quantity) are important for high-accuracy stress analysis. 
The Debye ring, observed by a two-dimensional detector, is observed as a ring-shaped X-ray intensity distribution with a finite width, and the intensity distribution takes the Gaussian-like form in the radial direction. 
For the residual stress analysis using the $cos \alpha$ method, the Debye ring deformation was analyzed as the shift of the peak of this Gaussian-like distribution. 
Thus, if the detector has insufficient spatial resolution, the accuracy of the peak position determination decreases. 
The intensity distribution is formed by the integrated value of the detector's output response for the collected charge quantity. 
Thus, if the detector has insufficient gain linearity, the intensity distribution of the diffraction ring will be distorted, which would affect the peak position determination. 
Therefore, for high-accuracy stress analysis, it is necessary to accurately determine the peak position from the X-ray image with high sharpness for the incident X-ray distribution and high linearity for the incident X-ray intensity. 

Spatial resolution performance is also required because of the shorter distance between the detector surface and the sample surface. 
For the application of residual stress measurements in industrial production, the main interest is in the surface layer of tens of \si{\micro\meter} (approximately \SI{100}{\micro\meter} or less) and X-rays below \SI{20}{k\electronvolt} are used because of the relatively large diffraction angle. 
For the measurement of the steel material, the X-ray energies of \SI{5.415}{k\electronvolt} (corresponding to the K$\alpha$ characteristic peak of the X-ray tube using the Cr target) or \SI{6.9}{k\electronvolt} (corresponding to the K$\alpha$ characteristic peak of the X-ray tube using the Co target) were mainly used. 
\SI{5.415}{k\electronvolt} is used for observing the diffraction peak with no overlap from the lattice plane (h, k, l) = (2, 1, 1) at $2\theta = 156.4$\si{\degree}, 
and \SI{6.9}{k\electronvolt} was used to observe the diffraction peak from the lattice plane (h, k, l) = (3, 1, 0) at $2\theta = 161.8$\si{\degree}. 
When using these X-ray energy conditions, the X-ray intensity loss caused by attenuation in air is not negligible. 
Thus, the camera length (CL), which is the distance between the detector surface and the sample surface, should be as small as possible. 
The spatial resolution of the detector is required to ensure accuracy in this small setup. 

The number of total input/output signal lines for the detector must be reduced as much as possible to reduce the size of the setup. 
To measure the entire shape of the Debye ring (more than 2 / 3) using a rectangular two-dimensional detector, more than two detector chips are necessary to cover the entire ring area while keeping the X-ray beam injection space. 
However, modern industrial products have complicated shapes and have several narrow parts; thus, the sensing unit including detectors must shrink when measurements are applied to such narrow parts. 
In this situation, there is not enough area to wire independent signal lines to each chip; thus, it is necessary to reduce the number of signal lines and to share the parts of the signal bus among all chips. 
To achieve this, the detector's output signal lines must have a switching function between the enable state and the disabled (high impedance) state. 

Furthermore, the detector should be able to withstand atmospheric conditions at room temperature (20--30 \si{\degreeCelsius}), because it is assumed that sufficient cooling mechanisms will not be provided in the assumed operating environment for residual stress measurement. 
Owing to the high thermal noise effect in this situation, multiple frames with short exposure times need to be stacked to obtain a sufficient exposure time. 
In the case of industrial measurements, such as the full inspection in the production line, the exposure time for a single measurement should be less than \SI{1}{s}. 
If this exposure time is met by adding \SI{100}{frames} of \SI{10}{ms} without increasing the time required, then a frame rate of \SI{100}{fps} or higher is required. 

\section{Overview of INTPIX4NA and Readout system}
\label{sec:i4na}

\subsection{INTPIX4NA specifications}

\begin{figure}[H]
\centering
\includegraphics[width=\linewidth]{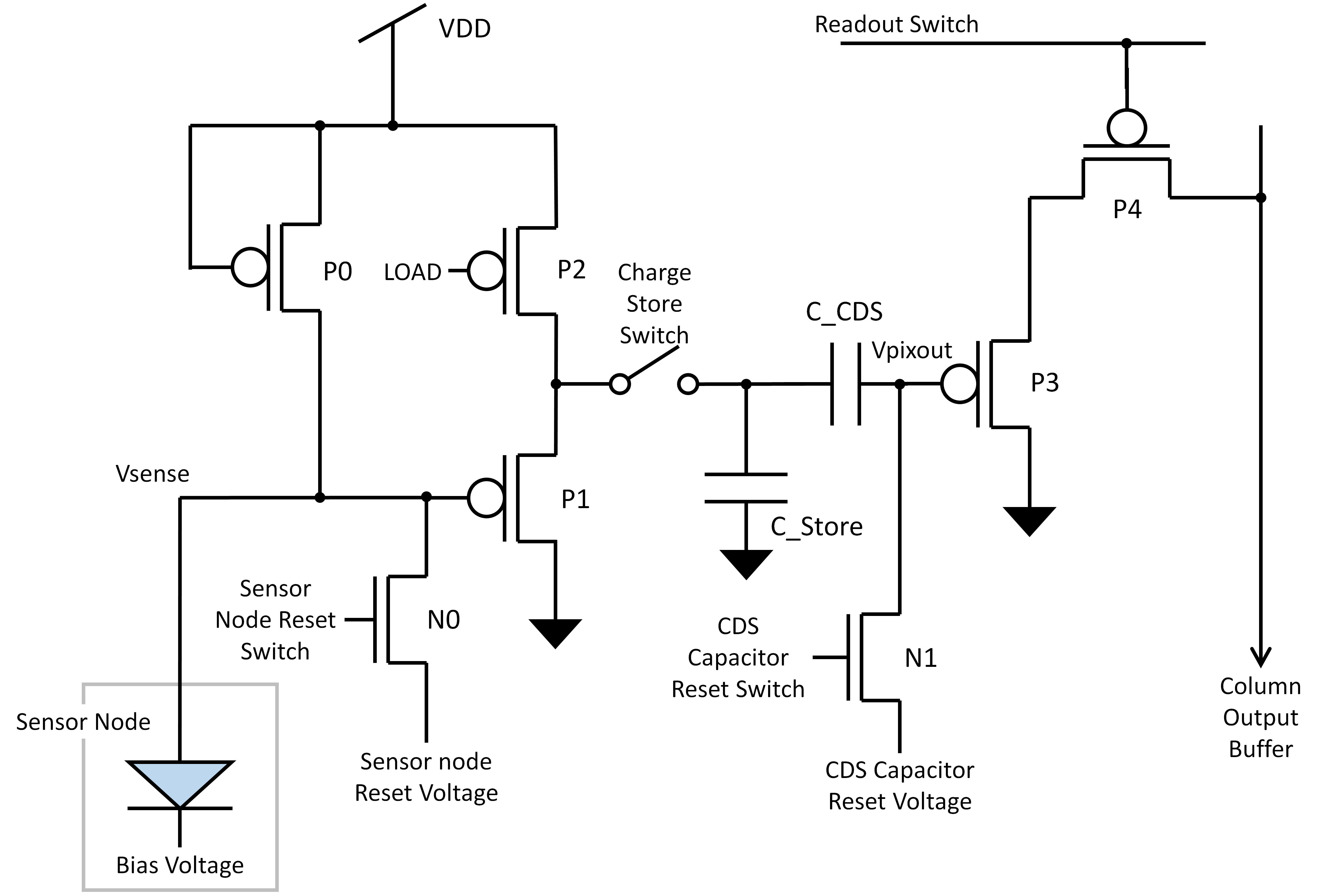}
\caption{INTPIX4NA In-Pixel circuit schematic.}
\label{fig:i4na_1}
\end{figure}

\begin{figure}[H]
\centering
\includegraphics[width=\linewidth]{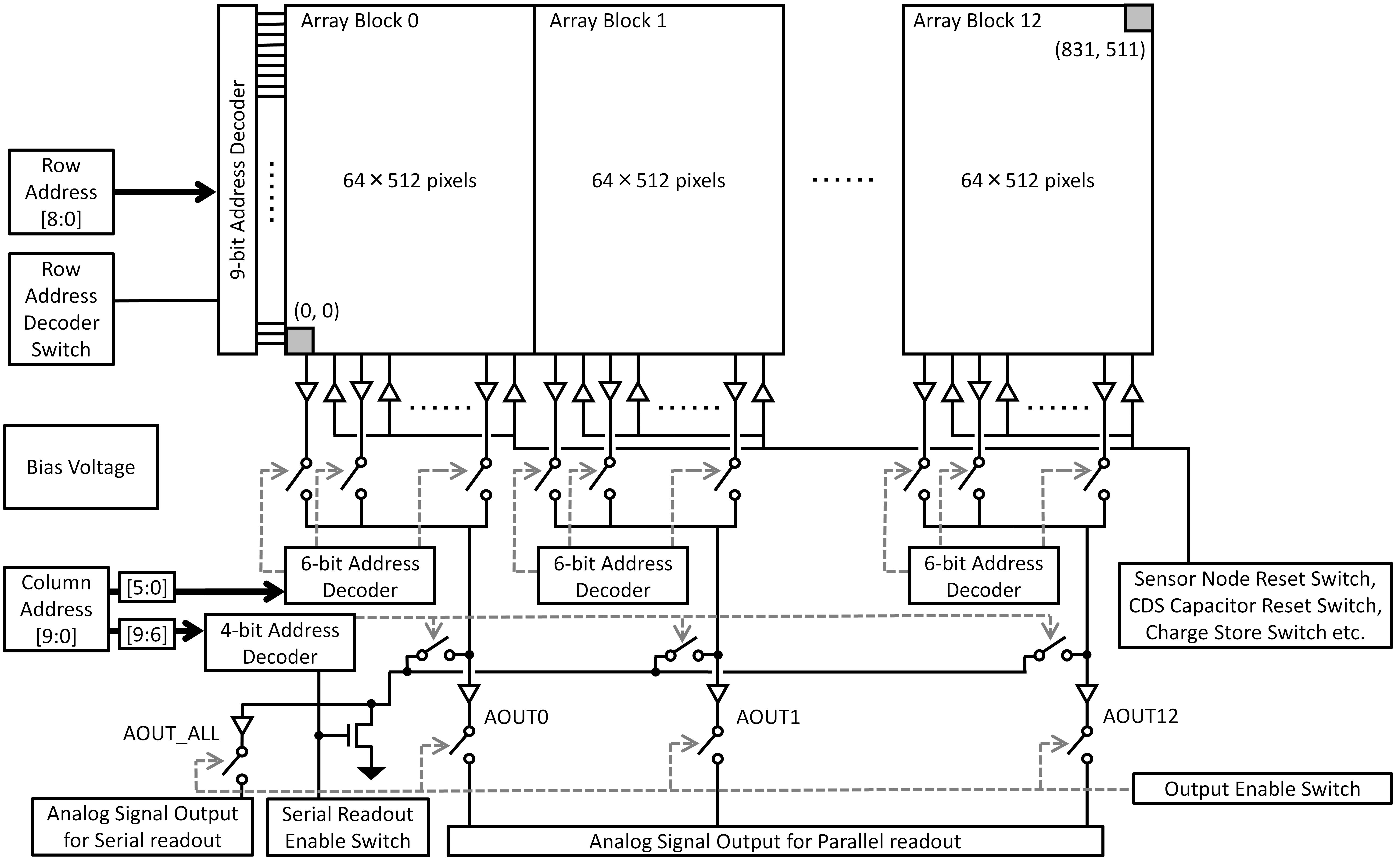}
\caption{INTPIX4NA whole pixel array and peripheral schematic.}
\label{fig:i4na_2}
\end{figure}

The INTPIX4NA is an integration-type SOIPIX detector developed based on INTPIX4.
The pixel size is 17 \si{\times} 17 \si{\micro\meter^2}, the number of pixels is 425,984 (832 column \si{\times} 512 row matrix), 
and the sensitive area is 14.1 \si{\times} 8.7 \si{mm^2}. 
The detector consists of 13 blocks (64 column \si{\times} 512 row pixel matrix contained in one block), and each block has an independent channel of analog output for parallel readout. 
Detailed design parameters are shown in table \ref{tab:detector_design}. 
An in-pixel circuit schematic is shown in figure \ref{fig:i4na_1}, and the whole pixel array and peripheral schematic are shown in figure \ref{fig:i4na_2}. 

\begin{table}[H]
 \centering
 \begin{tabular}{|c|c|}
 \hline
 Chip size & 15.4 \si{\times} 10.2 \si{mm^2} \\ \hline
 Sensitive area & 14.1 \si{\times} 8.7 \si{mm^2} \\ \hline
 Pixel matrix & \shortstack{\strut 832 columns \si{\times} 512 rows \\ (\SI{425984}{pixels})} \\ \hline
 Pixel size & 17 \si{\times} 17 \si{\micro\meter^2} \\ \hline
 Pixel gain & \SI{12.8}{\micro\volt/\elementarycharge} \\ \hline
 Shutter & Global Shutter \\ \hline
 Thickness of transistor layer & \SI{40}{nm} \\ \hline
 Thickness of buried oxide (BOX) layer & \SI{140}{nm} \\ \hline
 Thickness of sensor layer & \SI{300}{\micro\meter} (Typical) \\ \hline
 Sensor layer wafer & N type Floating Zone wafer \\ \hline
 \shortstack{\strut Back side \\ Aliminum deposition thickness} & \SI{150}{nm} (typical) \\ \hline
 Readout mode & \shortstack{\strut Full array one channel serial readout / \\ 13 channel parallel readout \\ (32,768 (64 columns \si{\times} 512 rows matrix) pixels/channel)} \\ \hline
 Number of signal lines & \shortstack{\strut 47 \\ (27 digital inputs, 4 analog inputs and 16 analog outputs) } \\ \hline
 Power Consumption & \SI{280}{mW} (Reset state) \\ \hline
 Other features & \shortstack{\strut In-Pixel Correlated Double \\ Sampling (CDS) circuit implemented.} \\ \hline
 \end{tabular}
 \caption{INTPIX4NA design parameters}
 \label{tab:detector_design}
\end{table}

The signal in the pixel is read out using the following procedure. 
Symbol names of part / signal follow figure \ref{fig:i4na_1}. 
At first, sensor node reset switch, CDS capacitor reset switch and charge store switch are closed. 
Vpixout is set to the voltage of the CDS capacitor reset voltage, and the sense node Vsence is set to the voltage of the sensor node reset voltage. 
The sensor node reset switch is opened before opening the CDS capacitor reset switch so as not to read out the reset noise from the sensor node reset switch.
Then exposure starts after opening the CDS capacitor reset switch. 
During exposure, the input signal is amplified by the source follower circuit consisting of P-type metal oxide semiconductor field effect transistors (MOSFETs) P1 and P2, and the potential Vpixout follows the potential of the amplified input voltage during exposure. 
Exposure stops after opening the charge store switch. 
Finally, Vpixout is read out when the readout switch is closed. 

The INTPIX4NA consists of 13 blocks, and each block consists of 64 \si{\times} 512 pixels (figure \ref{fig:i4na_2}). 
The signals of all the pixels are read out using two readout modes: the full array one channel serial readout mode and the 13 channels parallel readout mode. 
In the full array one channel serial readout mode, the signals are read out from one terminal using 6-bit (lower-order) and 4-bit (higher-order) column address decoders and a 9-bit row address decoder. 
In the 13 channels parallel readout mode, the signals are read out in 13 parallels using only the 6-bit column address decoder and a 9-bit row address decoder. 
These analog signals are converted to digital signals by analog-to-digital converters (ADCs) on a readout board \cite{seabas, ryunishiD, ryunishi3, ryunishi4}. 
The full array one channel serial readout mode can be used when the number of analog output signal lines should be reduced to one instead of allowing the readout speed to decrease. 
This readout mode can be used for a simple readout system that has only a single-channel ADC. 
The 13 channels parallel readout mode is the default readout mode of INTPIX4NA and can be used for a higher readout speed with a parallel output. 
In the residual stress measurement system, the 13 channels parallel readout mode was used. 

The basic specifications mentioned above were inherited from the original INTPIX4, and several specifications were implemented or applied in this INTPIX4NA. 
\begin{enumerate}
\setlength{\leftskip}{-10pt}
\setlength{\itemsep}{2pt}
\setlength{\parskip}{0pt}
\setlength{\itemindent}{0pt}
\setlength{\labelsep}{3pt}
\item Improving the circuit of ``AOUT buffer" (analog amplifier circuit placed in each blocks output) for the transmission of analog signal from in-pixel to readout for long distance transmission (less than 2 m). 
\item Improving the circuit of ``Column buffer" (analog amplifier placed in each column output) for the reduction of the settling time of analog output from more than \SI{320}{ns/pixel} to better than \SI{200}{ns/pixel} for higher framerates (more than \SI{150}{fps}). 
\item Implementation of the turn on / off function of the analog buffer to share one signal bus with several detectors. (Shown as ``Output Enable Switch" in figure \ref{fig:i4na_2}.)
When two or more INTPIX4NA chip analog outputs are connected to the same wiring, 
it is necessary to set the other outputs to a high-impedance state when one of the chips is in the output state to ensure normal signal propagation and prevent chip output destruction. 
This turn-on/off function will work as a switch for the analog buffer output state between the output mode and the high-impedance mode. 
\item Reduction of the sensor node reset time from \SI{5}{\micro\second} to less than \SI{1}{\micro\second} by reducing resistyvity of lines. 
\item Changing the field oxide process in SOI chip manufacturing from local oxidation of silicon (LOCOS) to shallow trench isolation (STI) for pixel array yield improvement. 
The field oxide process of chip manufacturing involves a baking process (800--1,200 \si{\degreeCelsius}). 
The floating zone wafer, used for the sensing part of SOIPIX, is relatively sensitive to thermal stress, and slips are formed during the baking process in the LOCOS process. 
The STI process uses a lower temperature for the baking process (800--1,000 \si{\degreeCelsius}); thus, slips are reduced from past SOIPIX production using the LOCOS process. 
\item Changing the silicon wafer's resistivity of the Si substrate for the sensing part from \SI{2} to 10.1--12.1 \si{k \ohm \cdot cm} to reduce the detector bias voltage. 
\end{enumerate}
The improvement of ``AOUT buffer" and ``Column buffer" circuits were expected to affect the response characteristics of the signals. 
The changes in the characteristics of silicon wafers and semiconductor processes may affect the spatial resolution characteristics. 
Thus, we planned to test the characteristics of spatial resolution, energy resolution, and gain linearity. 
In addition, the settling time in the practical setup was also tested. 

\subsection{INTPIX4NA setup with Readout system}

\begin{figure}[H]
\centering
\includegraphics[width=\linewidth]{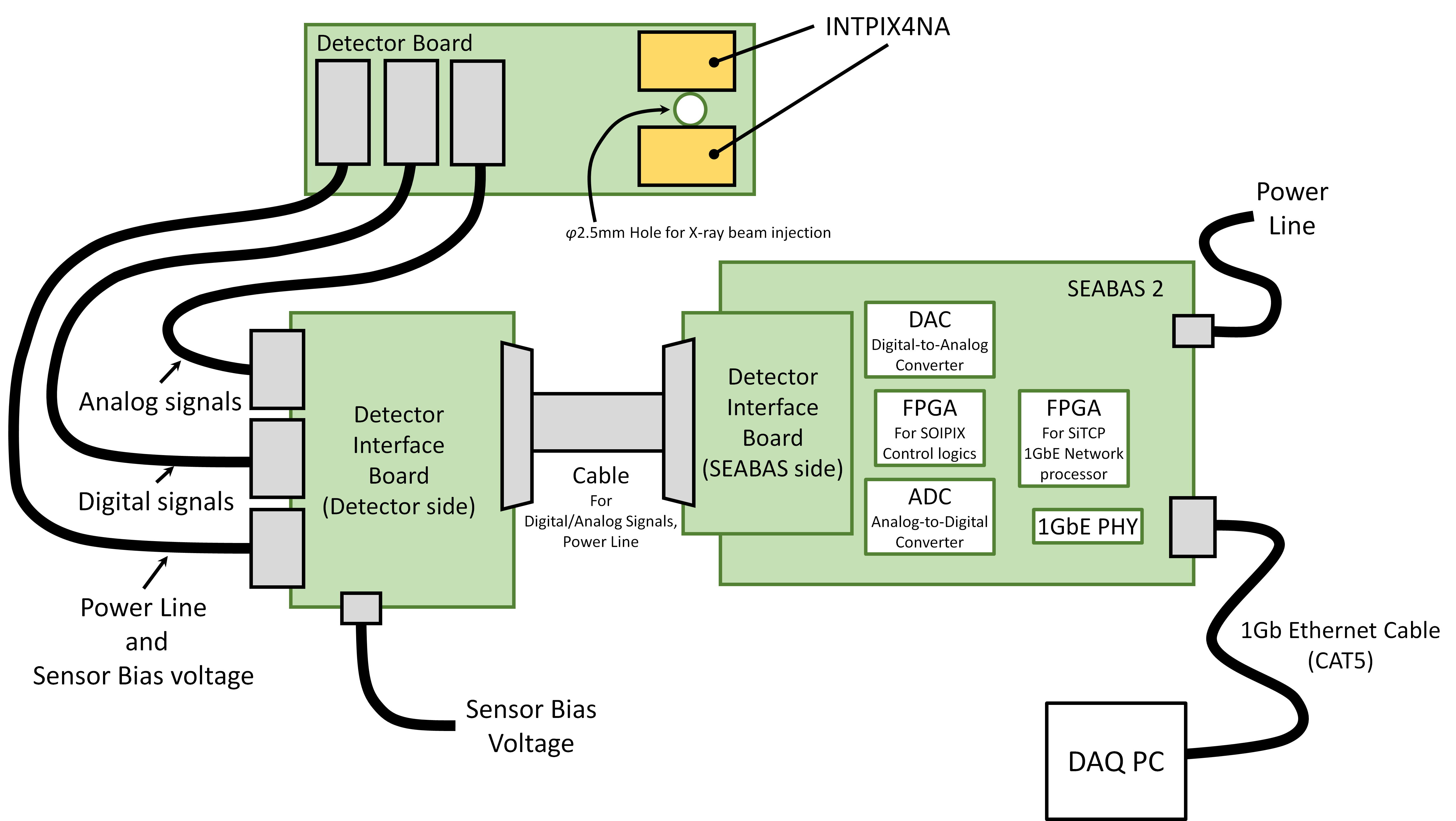}
\caption{INTPIX4NA and SEABAS2 connection schematic.}
\label{fig:seabas2andintpix4na}
\end{figure}

INTPIX4NA signals are read out by the data acquisition (DAQ) system based on the SEABAS2 (Soi EvAluation BoArd with Sitcp 2) \cite{seabas, ryunishiD, ryunishi3, ryunishi4} readout board. 
SEABAS2 has a 16-channel analog-to-digital converter, a four-channel digital-to-analog converter, two field-programmable gate arrays (FPGAs), and a 1 gigabit Ethernet I/F (1 GbE). 
The entire setup of the INTPIX4NA and SEABAS2 connections is shown in figure \ref{fig:seabas2andintpix4na}. 
For the residual stress measurement, the DAQ system consists of a SEABAS2 board, the sensing board shown as ``Detector board" in figure \ref{fig:seabas2andintpix4na}, two mezzanine boards shown as ``Detector Interface board (Detector side)" and ``Detector Interface board (SEABAS2 side)" in figure \\\\ ref{fig:seabas2andintpix4na}, and DAQ PC (Standard PC/AT compatible machine). 
In this setup, two INTPIX4NA chips are mounted on the detector board. 
The detector board has a hole at the center between the two chips for the X-ray beam to pass through. 
The outside dimensions of the detector board are 105 \si{\times} 33 \si{mm^2}, the edge-to-edge distance between the chips is typically \SI{3}{mm} (from the dicing edge of the chip) and 3.639 mm (from the edge of the sensitive area), and the diameter of the hole for the X-ray beam is \si{\phi}\SI{2.5}{mm}. 
The detector board is connected to SEABAS2 via two mezzanine boards. 
The mezzanine boards closer to the detector board are called ``Detector Interface board (Detector side)" and the other is ``Detector Interface board (SEABAS2 side)". 
The detector board and detector interface board (detector side) are connected by three cables. 
A total of 21 poles (1 no contact) coaxial cable for analog signals, 31 poles (1 no contact) coaxial cable for digital signals, and 5 poles parallel cable for power lines/sensor bias voltage line. 
Each cable length is \SI{1}{m} (typical). 
Two detector interface boards are connected by one cable, 68 Parallel poles for analog and digital signals, and several power lines. 
The cable length is 0.2 -- 2 m. 
The detector interface board (SEABAS2 side) and SEABAS2 were connected via an IEEE P-1386 mezzanine connector. 
SEABAS2 is connected to the DAQ computer via 1 gigabit Ethernet, and all data/commands are transferred by TCP and UDP protocols. 

\section{Experiments}
\label{sec:exp}

We performed the following experiments at the synchrotron beamlines of the Photon Factory (KEK) BL-14A and BL-14B. 
The experiments were performed in the air at room temperature (approximately 25--27 \si{\degreeCelsius}) using monochromatic X-rays of \SI{5.415}, \SI{8}, and \SI{12}{k\electronvolt}.

\subsection{Spacial resolution characteristics}
\label{sec:exp1}

\begin{figure}[H]
\centering
\includegraphics[width=\linewidth]{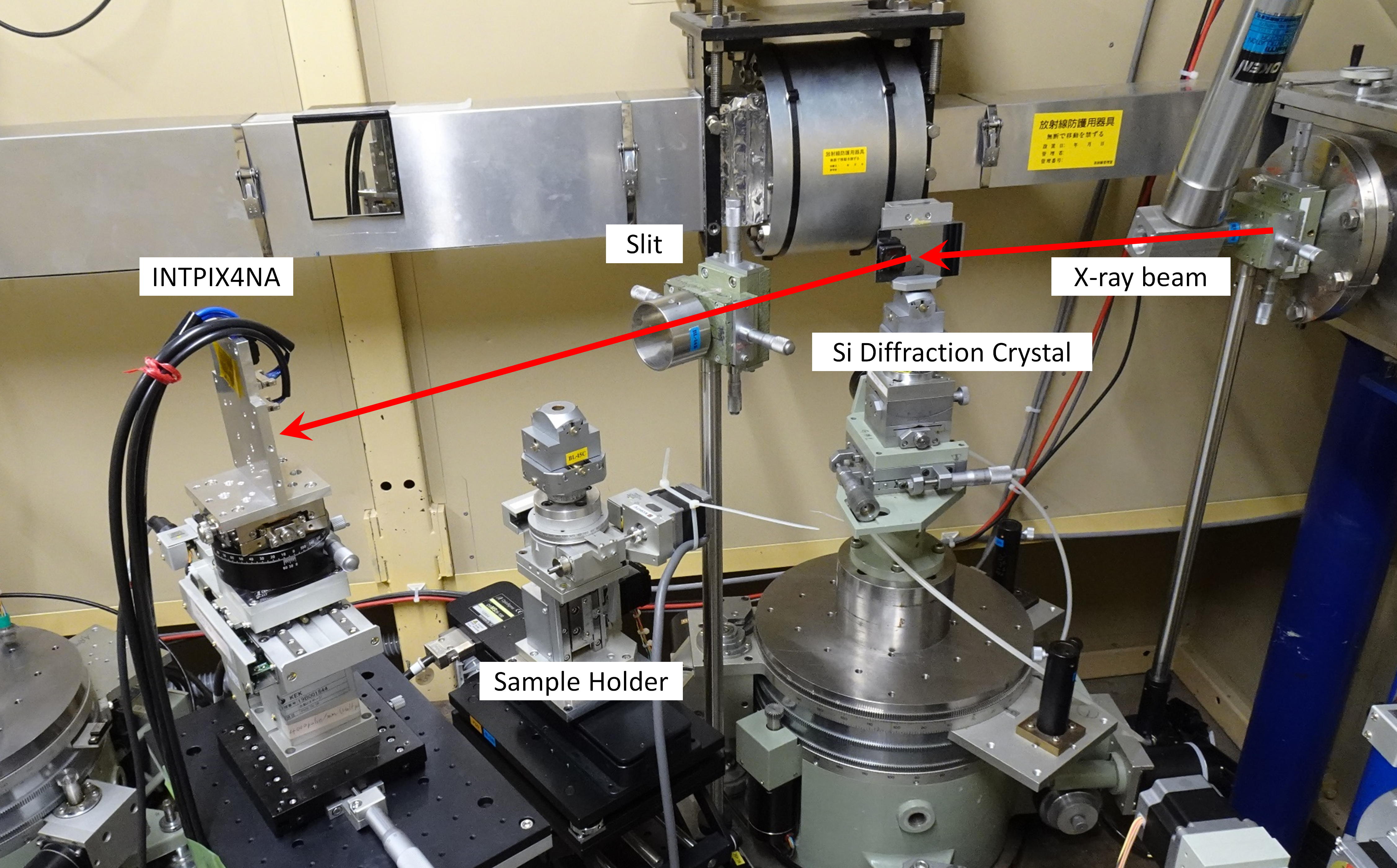}
\caption{Spacial resolution measurement setup at BL-14B.}
\label{fig:exp_14b}
\end{figure}

The spatial resolution measurement was performed at BL-14B using a \SI{12}{k\electronvolt} monochromatic X-ray beam. 
An X-ray beam was injected from the backside (opposite to the circuit layer) of the detector. 
The sample to generate slanted-edge was \si{\phi}\SI{0.84}{mm} Nickel-plated needle. 
Detector operation parameters were shown in table \ref{tab:exp_setup_detail_soi_1}.
Figure \ref{fig:exp_14b} is the whole setup for this measurement (sample was not placed on the sample holder in this figure). 

\begin{table}[H]
 \centering
 \begin{tabular}{|c|c|}
 \hline
 X-ray Energy & \SI{12}{k\electronvolt} monochromatic \\ \hline
 Sensor bias voltage & +\SI{250}{\volt} \\ \hline
 Sensor temperature & Room temperature (approximately 25--27 \si{\degreeCelsius}) \\ \hline
 Exposure time & \SI{250}{ms} (\SI{250}{\micro\second/frame} \si{\times} \SI{1000}{frames}) \\ \hline
 \shortstack{\strut Scan time \\ (Settling wait time \\for analog output)} & \SI{320}{ns/pixel} \\ \hline
 Sensor node reset & \SI{1}{\micro\second/frame} with \SI{300}{mV} ref. voltage \\ \hline
 CDS reset & \SI{2}{\micro\second/frame} with \SI{350}{mV} ref. voltage \\ \hline
 Readout board & SEABAS 2 \cite{seabas, ryunishiD, ryunishi3, ryunishi4} \\ \hline
 \end{tabular}
\caption{INTPIX4NA detector operation parameters for spatial resolution measurement.}
\label{tab:exp_setup_detail_soi_1}
\end{table}

The spatial resolution is evaluated by measuring the modulation transfer function (MTF). 
The MTF is the sharpness index, which is defined as the absolute of the optical transfer function (OTF, signal transfer characteristics in the spatial frequency domain). 
The MTF is calculated from the edge spread function (ESF) of the composite slanted-edge profile, 
the line spread function (LSF) is defined as the derivative of the ESF, 
and the OTF is defined as the Fourier transformation of the LSF. 
In this study, the slanted-edge method \cite{mtf} was used for MTF measurement. 
To measure the MTF of a digital imaging system, such as a two-dimensional pixel detector, the composite edge spread function (ESF) is used to avoid aliasing errors. 
This composite ESF is calculated from edge profiles with edge positions gradually shifted from to the pixel boundary orthogonal to the scanning direction of the profile. 
The MTF obtained using this composite ESF is referred to as the presampled MTF. 
The slanted-edge method was used to obtain edge profiles to calculate the composite ESF. 
In this slanted-edge method, the image taken with the edge line tilted several degrees from the pixel boundary orthogonal to the scanning direction of the profile is used. 
The composite ESF is obtained as a composite profile of several edge profiles aligned by the edge line position. 
In this measurement, the X-ray image of the slanted edge generated by the shadow of the metal sample and the X-ray image of the beam profile were taken first. 
Then the slanted-edge image was normalized by the beam profile image. 
This normalized slanted-edge image was the sampling source of the ESF from the normalized slanted-edge. 

\begin{figure}[H]
\centering
\includegraphics[width=\linewidth]{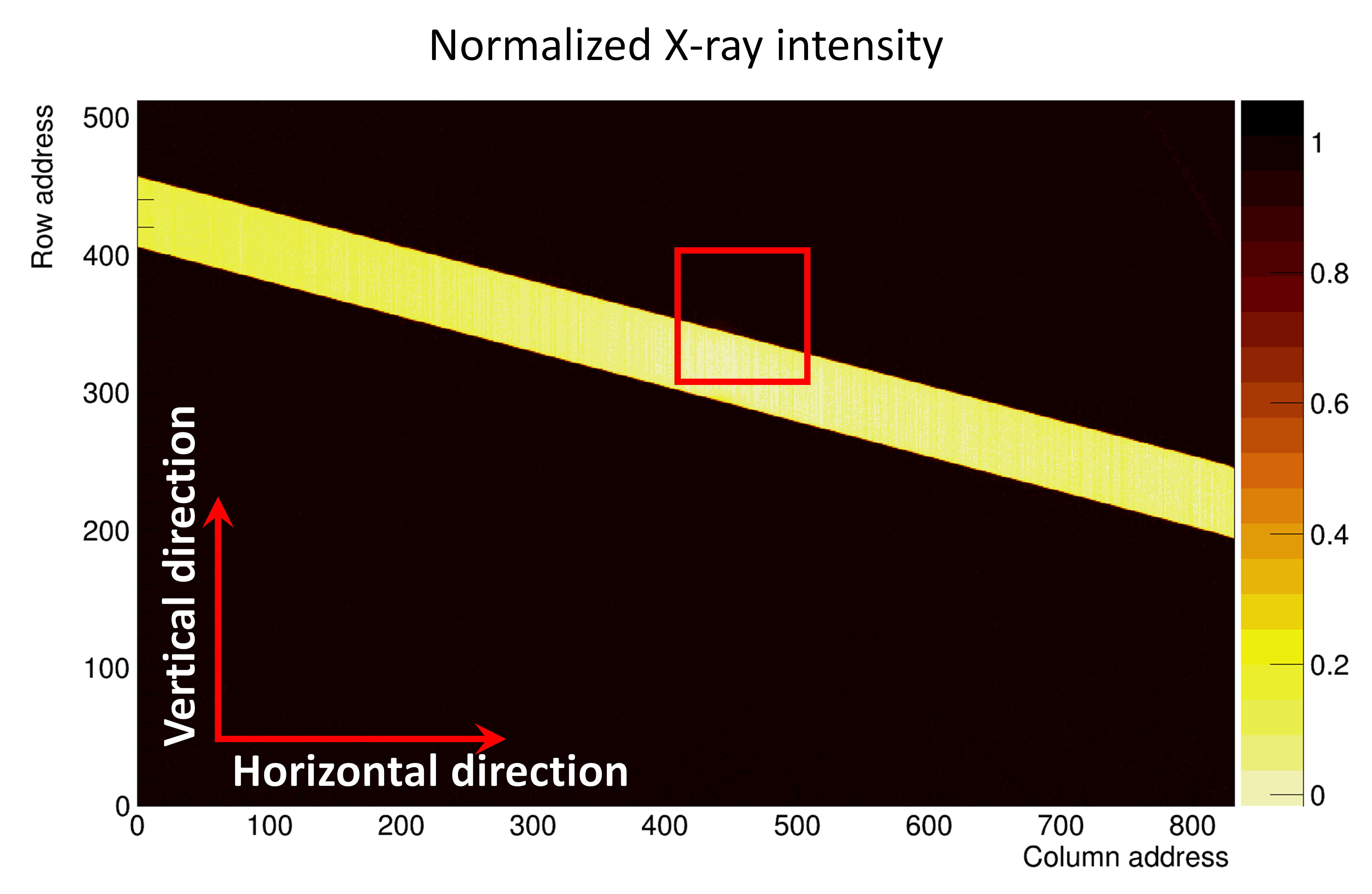}
\caption{Normalized X-ray slanted-edge image of Nickel-plated needle. The frame shown by the red line is used for the vertical direction ESF calculation.}
\label{fig:exp_14b_r1}
\end{figure}

\begin{figure}[H]
\centering
\includegraphics[width=\linewidth]{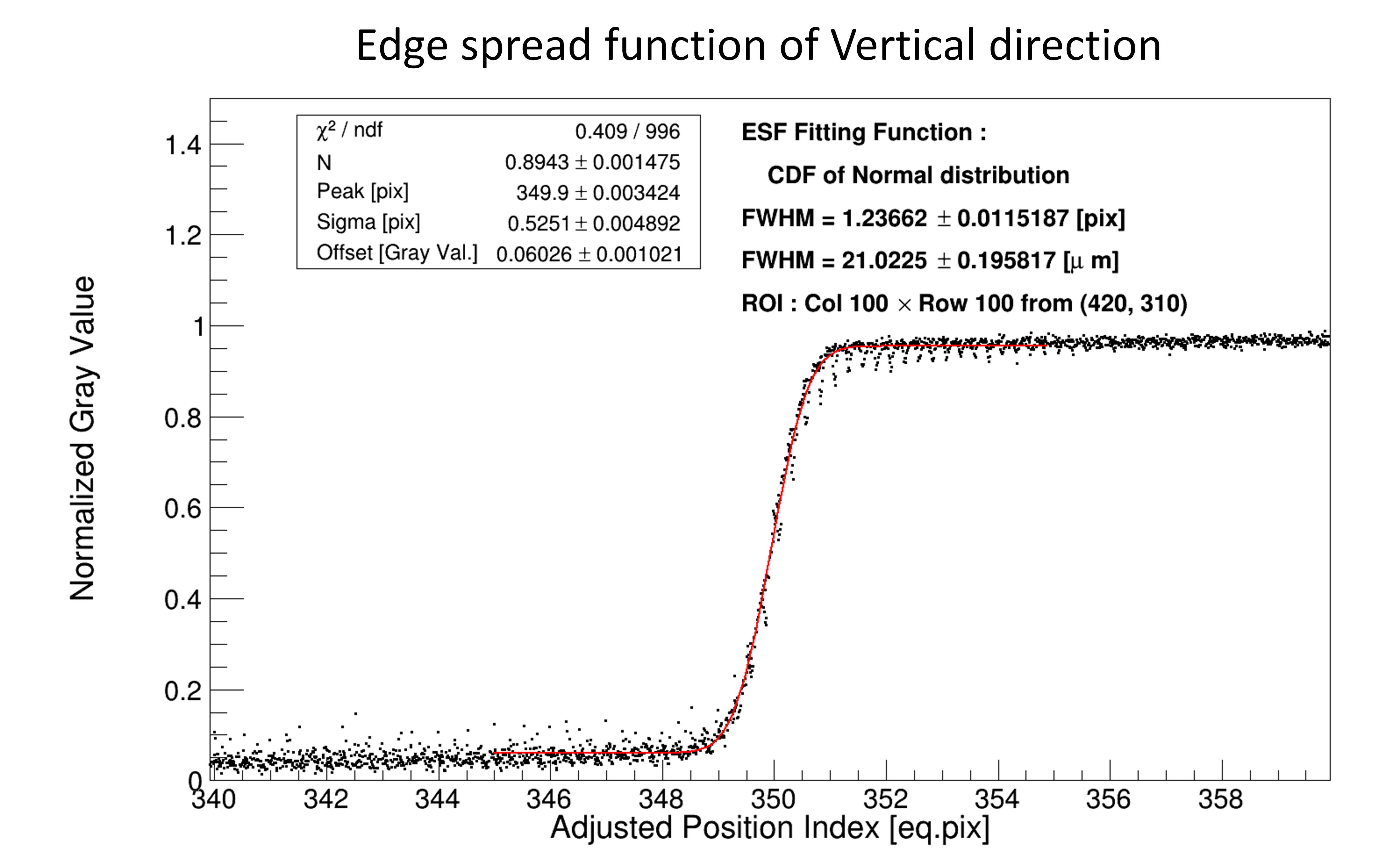}
\caption{Vertical direction ESF plot.}
\label{fig:exp_14b_r2}
\end{figure}

\begin{figure}[H]
\centering
\includegraphics[width=\linewidth]{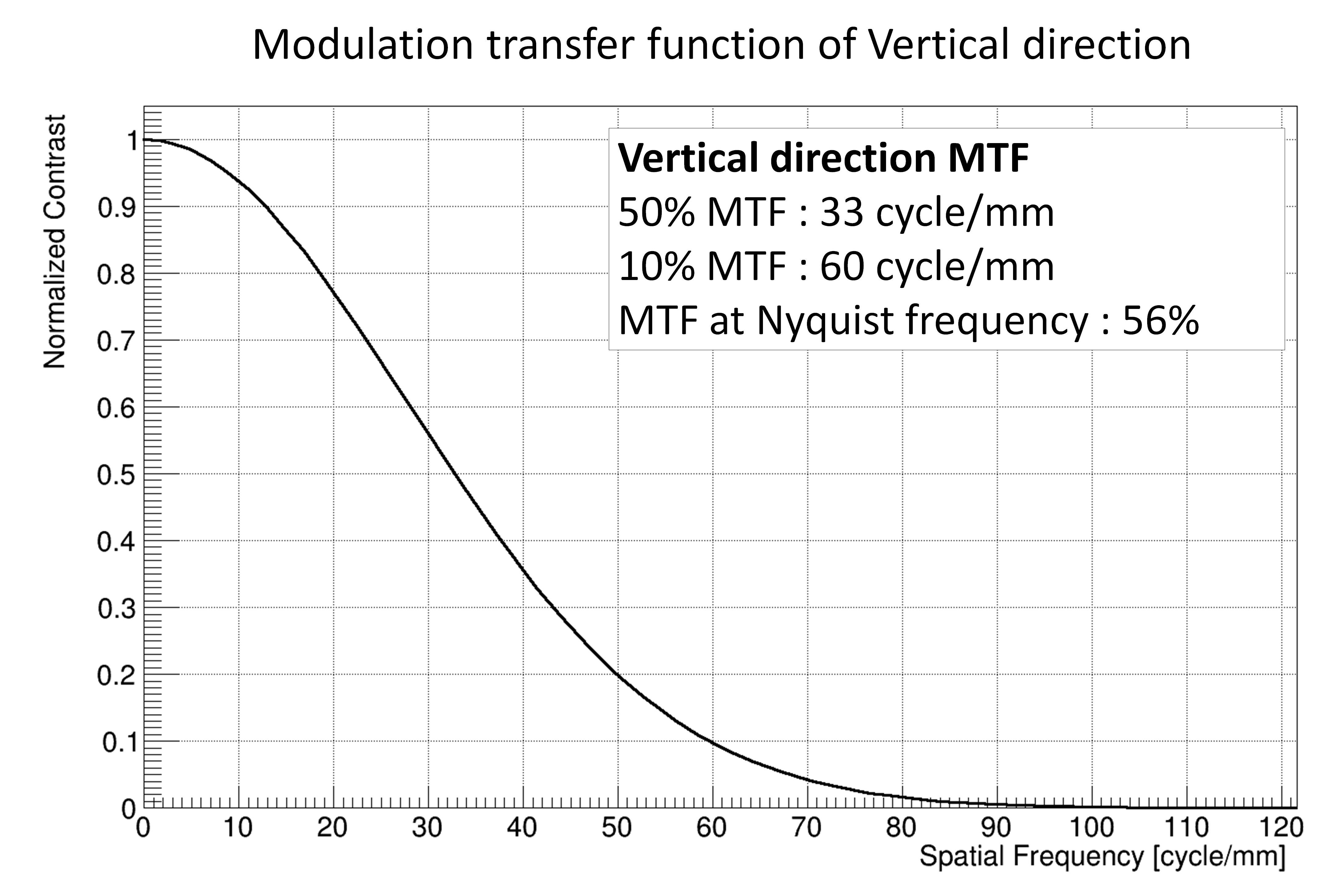}
\caption{Vertical direction MTF plot.}
\label{fig:exp_14b_r3}
\end{figure}

\begin{figure}[H]
\centering
\includegraphics[width=\linewidth]{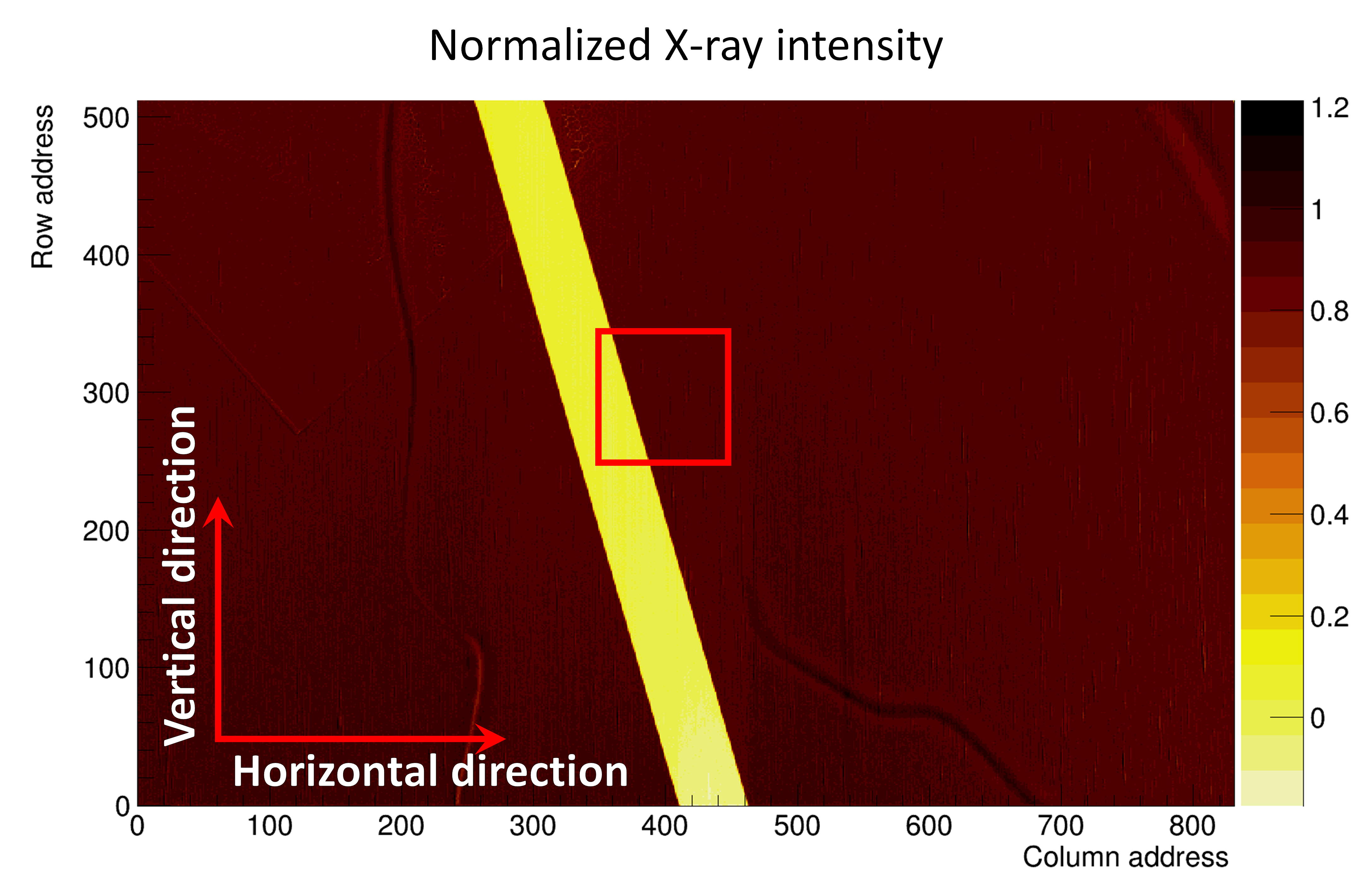}
\caption{Normalized X-ray slanted-edge image of Nickel-plated needle. The frame shown by the red line is used for the horizontal direction ESF calculation.}
\label{fig:exp_14b_r4}
\end{figure}

\begin{figure}[H]
\centering
\includegraphics[width=\linewidth]{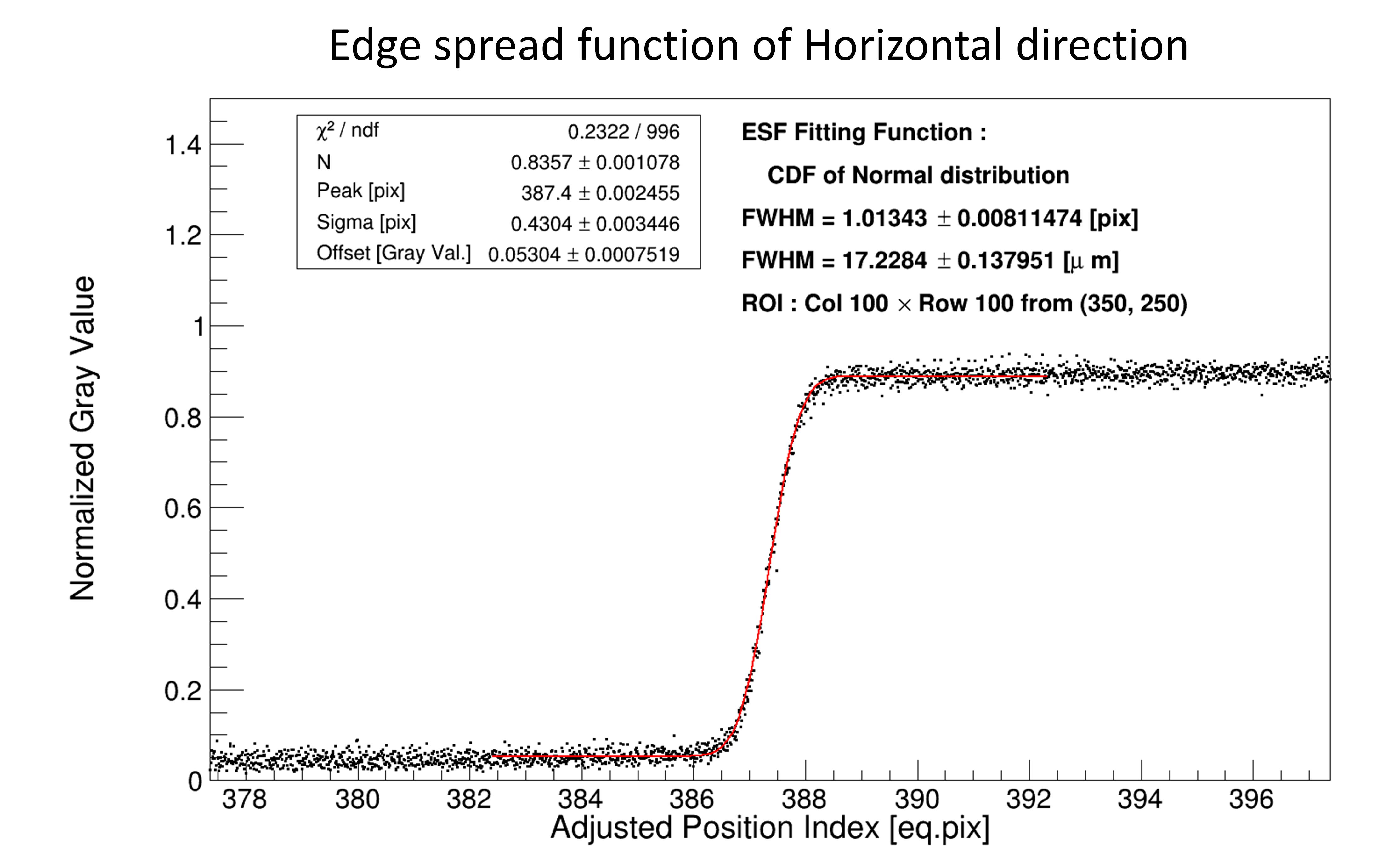}
\caption{Horizontal direction ESF plot.}
\label{fig:exp_14b_r5}
\end{figure}

\begin{figure}[H]
\centering
\includegraphics[width=\linewidth]{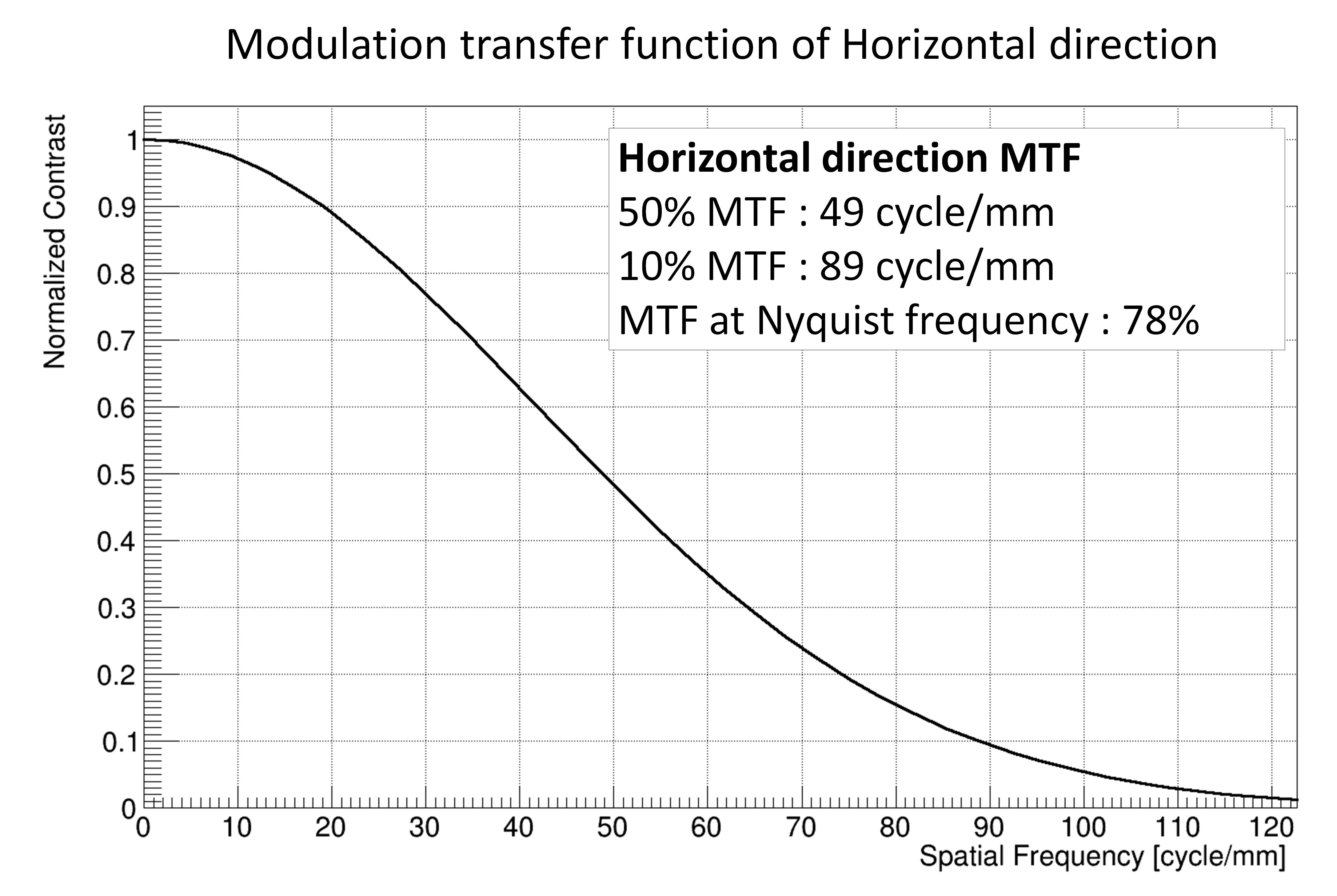}
\caption{Horizontal direction MTF plot.}
\label{fig:exp_14b_r6}
\end{figure}

In the results of the spatial resolution measurement, short axis direction is labelled as ``Vertical" and the long axis direction is labelled as ``Horizontal". 
The results of the MTF in the vertical direction are shown in figure \ref{fig:exp_14b_r1}, \ref{fig:exp_14b_r2}, and \ref{fig:exp_14b_r3}. 
Figure \ref{fig:exp_14b_r1} is the slanted-edge X-ray image normalized by the beam profile image. 
Figure \ref{fig:exp_14b_r2} is the ESF of the slanted-edge, and parameters of LSF are shown in the top of the figure.
Figure \ref{fig:exp_14b_r3} is the final result of the MTF. 
Figure \ref{fig:exp_14b_r4}, \ref{fig:exp_14b_r5}, \ref{fig:exp_14b_r6} are the same for the MTF in the horizontal direction. 
INTPIX4NA showed more than a 50\% at the Nyquist frequency (\SI{29.4}{cycle/mm}) in both the vertical and horizontal directions. 
In addition, 50\% and 10\% MTF were beyond the Nyquist frequency in both the vertical and horizontal directions. 
Thus, this detector's spatial resolution performance reached the limitation defined by pixel size. 
These resolution characteristics are superior to those of the indirect conversion-type device of the X-ray sCMOS camera \cite{xscmos1, xscmos2} below \SI{20}{k\electronvolt}. 
These are also superior to those of recent direct conversion-type devices \cite{percival}. 

\subsection{Energy resolution and Gain linearity}
\label{sec:exp2}
In the X-ray residual stress measurement, the detector was operated with a low-energy X-ray beam under room temperature (20--30 \si{\degreeCelsius}) conditions. 
Thus, the detector must have a signal-to-noise ratio sufficient to separate the signal of a single photon from the high thermal noise at the X-ray energy to be used. 
Furthermore, the gain linearity, the linearity of the output response for the collected charge quantity, is important for improving the accuracy of the residual stress value. 
The intensity peak position of the diffraction ring was used to calculate the residual stress. 
This intensity distribution is formed by the integrated value of the detector's output response for the collected charge quantity. 
Thus, if the detector has insufficient gain linearity, the intensity distribution of the diffraction ring will be distorted, and the accuracy of the peak position determination will decrease. 
To confirm these performances, we evaluated the energy resolution using X-rays in the energy range used for X-ray residual stress measurement and analyzed the gain linearity from the energy resolution results. 

\begin{figure}[H]
\centering
\includegraphics[width=\linewidth]{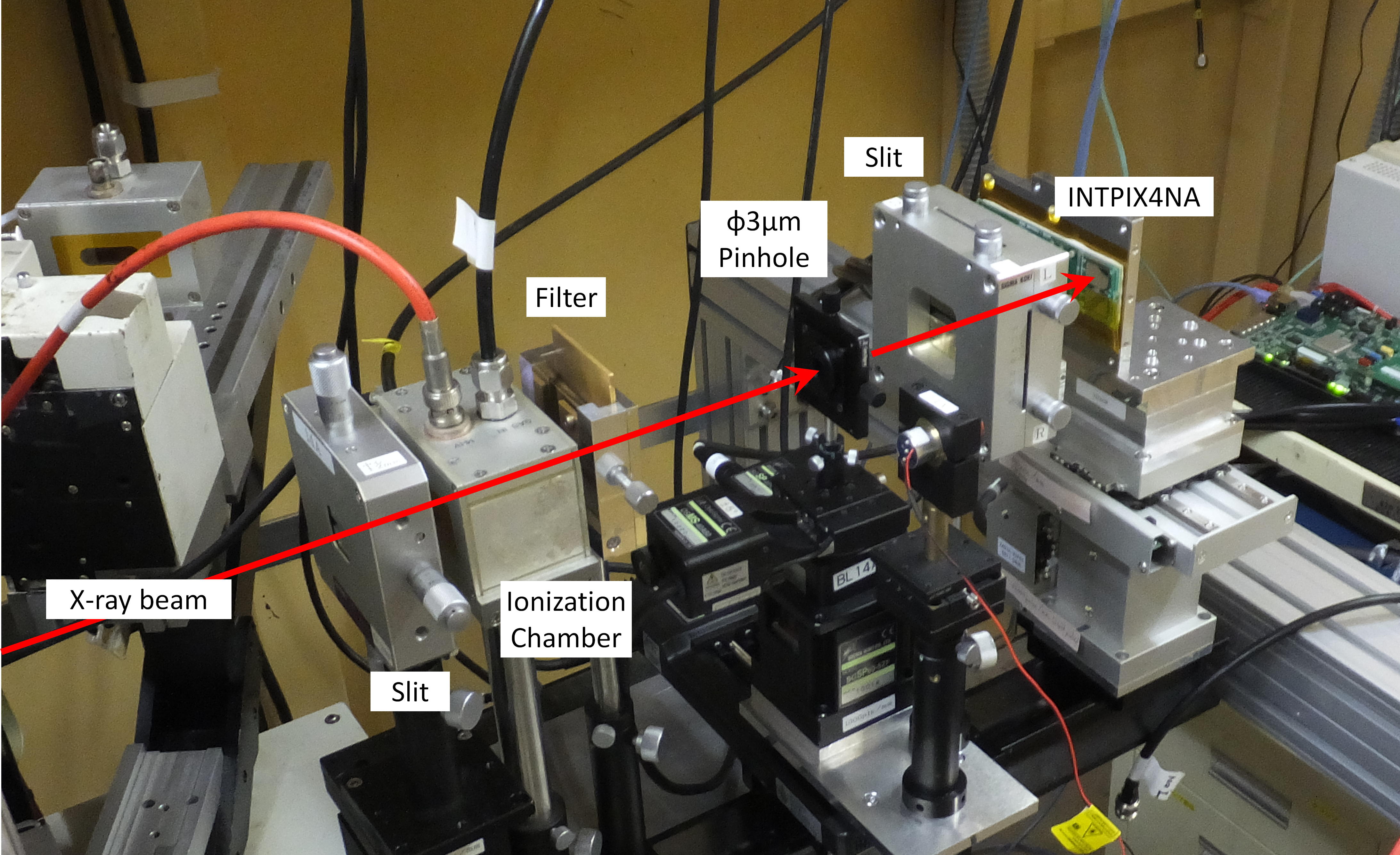}
\caption{Energy resolution and Gain linearity measurement setup at BL-14A.}
\label{fig:exp_14a}
\end{figure}

\begin{figure}[H]
\centering
\includegraphics[width=\linewidth]{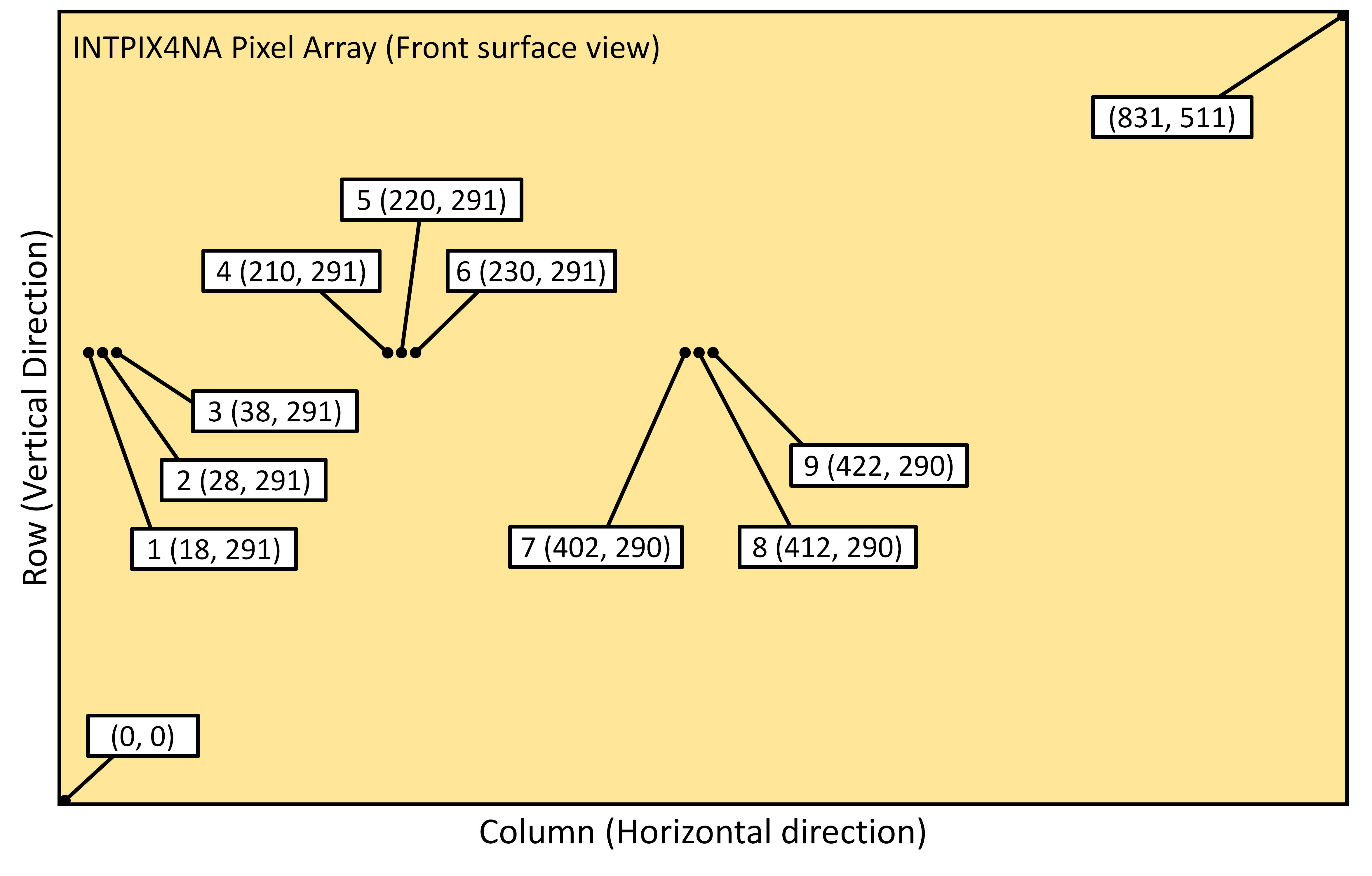}
\caption{Map of target pixels for energy resolution and gain linearity measurements.}
\label{fig:targetmap}
\end{figure}

\begin{figure}[H]
\centering
\includegraphics[width=\linewidth]{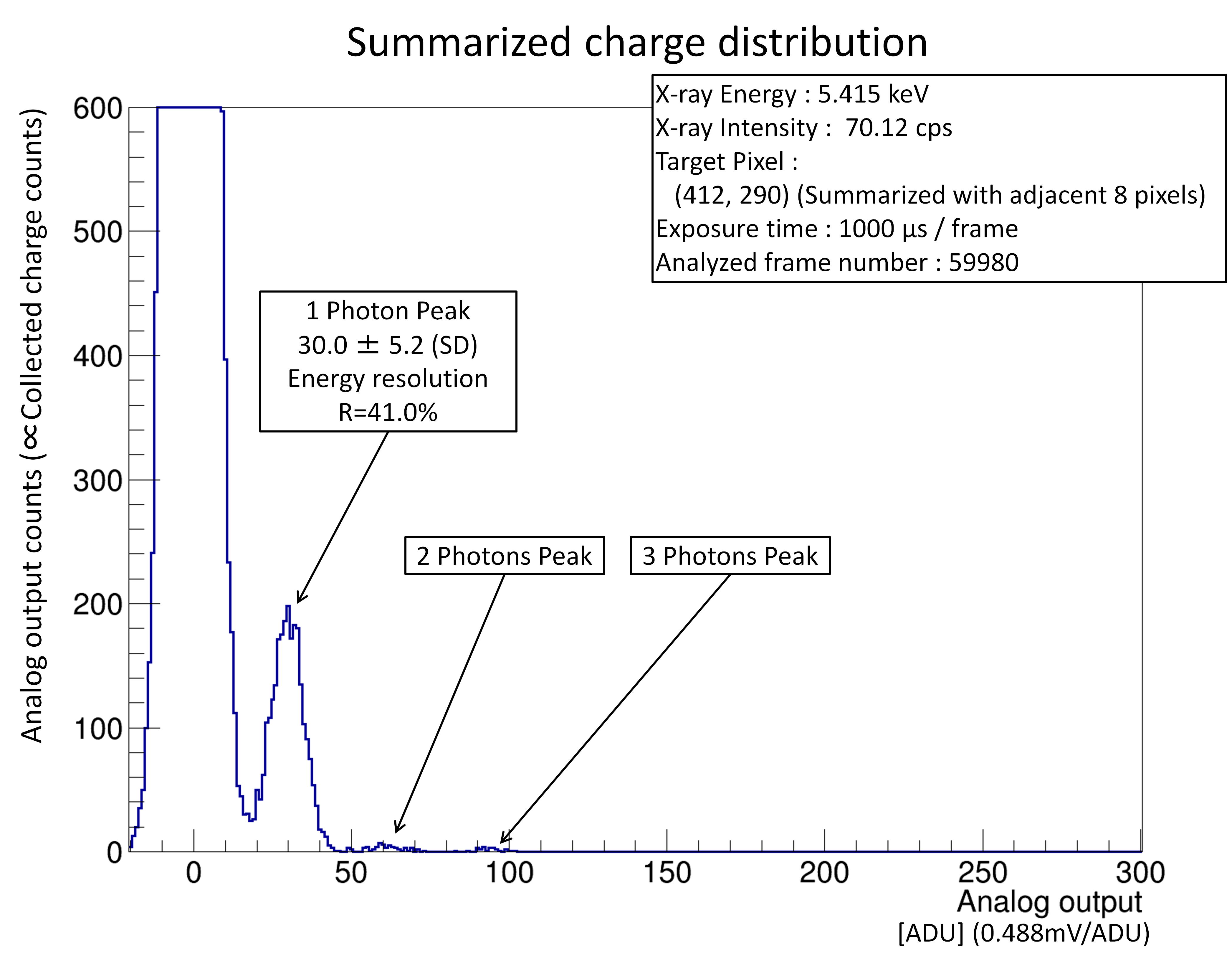}
\caption{Distribution of summarized charge of the target pixel No.8 at E = \SI{5.415}{k\electronvolt}.}
\label{fig:energypeak_no8_5kev}
\end{figure}

\begin{figure}[H]
\centering
\includegraphics[width=\linewidth]{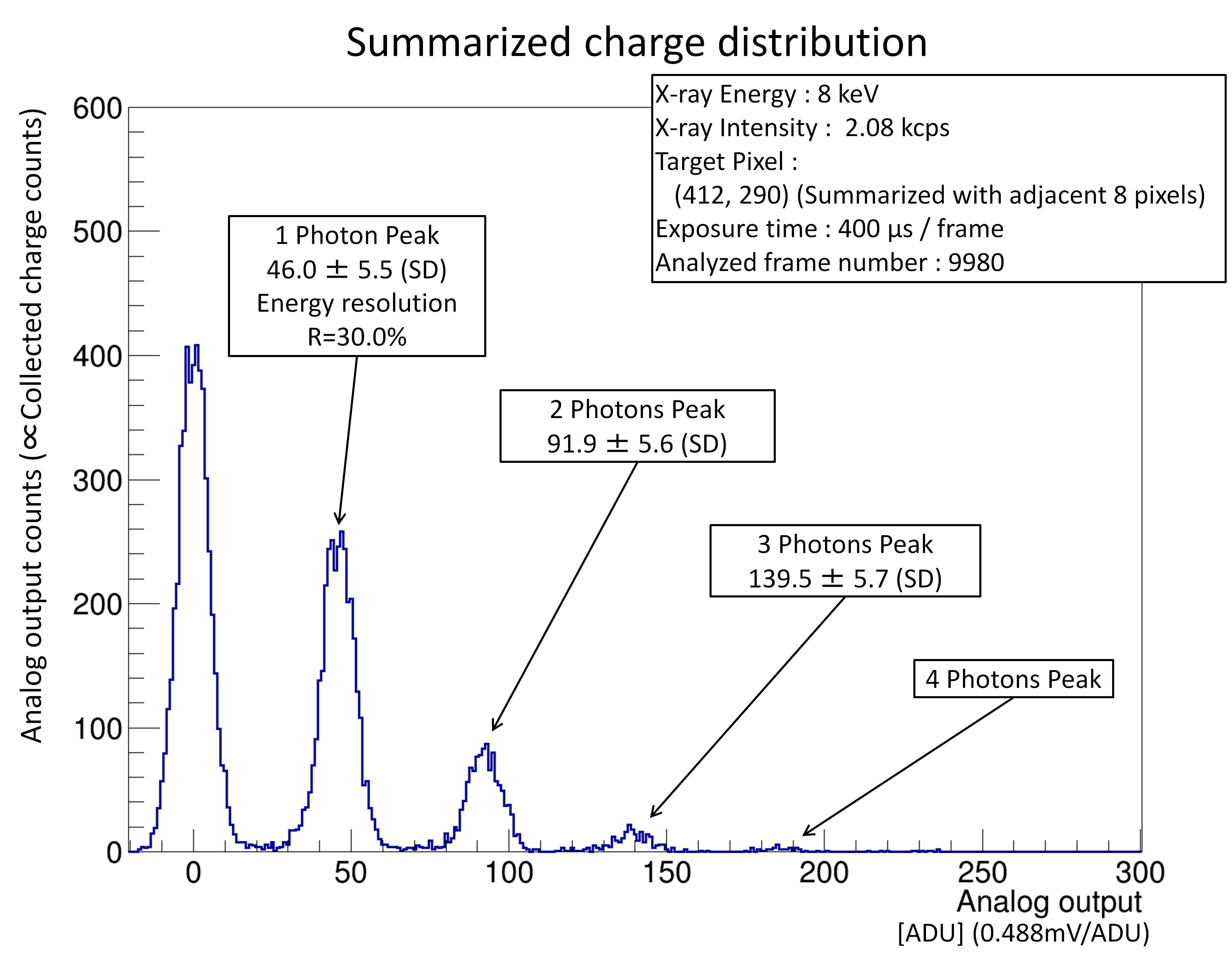}
\caption{Distribution of summarized charge of the target pixel No.8 at E = \SI{8}{k\electronvolt}.}
\label{fig:energypeak_no8_8kev}
\end{figure}

\begin{figure}[H]
\centering
\includegraphics[width=\linewidth]{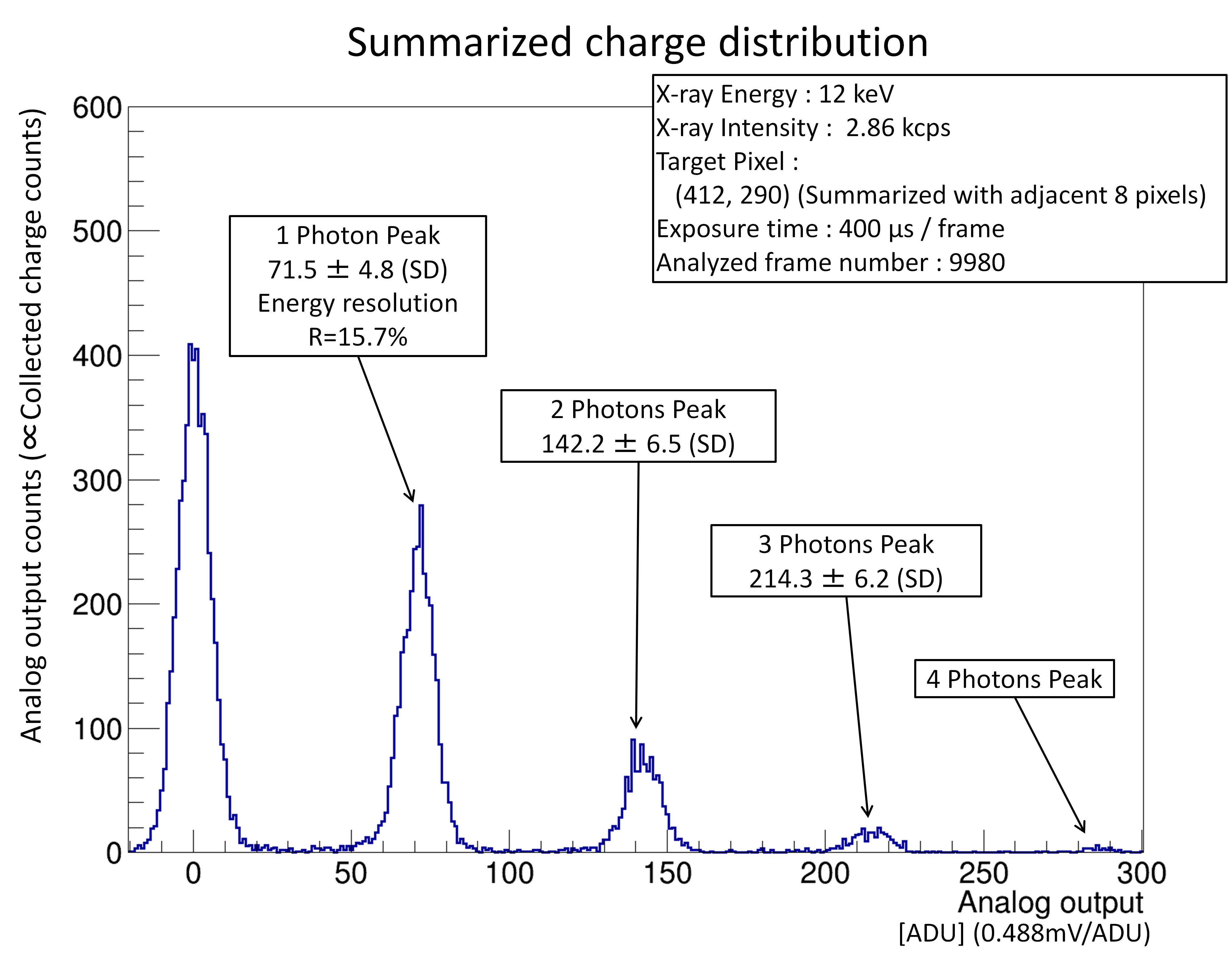}
\caption{Distribution of summarized charge of the target pixel No.8 at E = \SI{12}{k\electronvolt}.}
\label{fig:energypeak_no8_12kev}
\end{figure}

The energy resolution was measured at BL-14A using \SI{5.415}, \SI{8}, and \SI{12}{k\electronvolt} monochromatic X-ray beams. 
The X-ray beam diameter was reduced by 3um pinholes and injected from the backside to target pixels on the detector pixel array. 
And the analog signal, proportional to the collected charge quantity, was read out from the target pixels after every exposure time. 
In these measurements, the beam diameter at the detector surface is \si{\phi}\SI{10}{\micro\meter} (FWHM), and the charge diffusion in the sensor layer was \textasciitilde \si{\pm{}}\SI{10}{\micro\meter}. 
In such a situation, the total distribution area of the charge generated from the X-ray photon is larger than the actual pixel size. 
Therefore, in these measurements, the target pixel's analog signal output was added to the output of the neighbor 8 pixels. 
Thus, the energy resolution was analyzed as a virtual 51 \si{\times} 51 \si{\micro\meter^2} pixel. 
Detector operation parameters were shown in table \ref{tab:exp_setup_detail_soi_2}.
Figure \ref{fig:exp_14a} is the whole setup of this measurement. 
In these measurements, nine pixels (with neighbors) were chosen as the target. 
The target pixels are listed in table \ref{tab:exp_targetpix} and mapped in figure \ref{fig:targetmap}. 
The pixel identification numbers in figures \ref{fig:energypeak_no8_5kev}, \ref{fig:energypeak_no8_8kev}, and \ref{fig:energypeak_no8_12kev}, and table \ref{tab:exp_energypeak_5kev}, \ref{tab:exp_energypeak_8kev}, and \ref{tab:exp_energypeak_12kev} correspond to the numbers listed in table \ref{tab:exp_targetpix}.

In this study, the energy resolution is defined as the value R obtained from the peak height of the charge distribution generated by one-photon injection ($\mathrm{{H}_{peak}}$) and the full width at half maximum (FWHM) of the peak using the following equation: $\mathrm{R = FWHM / {H}_{peak} \times 100 [\%]}$.
$\mathrm{{H}_{peak}}$ and the FWHM can be obtained from the distribution of charges from each pixel. 
The gain linearity can be analyzed using the position of each peak in the energy resolution results described above and the estimated charge count corresponding to the peak position. 
The number of collected charges for each peak's height can be estimated from the average energy of electron-hole pair production (known as \SI{3.6}{\electronvolt} for silicon) and related X-ray energy. 
Thus, the gain per elementary charge is obtained from the relationship between the peak of the signal and the estimated number of charges related to the peak. 

\begin{table}[H]
 \centering
 \begin{tabular}{|c|c|}
 \hline
 X-ray Energy & 5.415, 8 and \SI{12}{k\electronvolt} monochromatic \\ \hline
 Sensor bias voltage & +\SI{250}{\volt} \\ \hline
 Sensor temperature & Room temperature (approximately 25--27 \si{\degreeCelsius}) \\ \hline
 Exposure time & \shortstack{\strut \SI{400}{\micro\second/frame} \si{\times} \SI{10000}{frames} for \SI{8}{k\electronvolt} and \SI{12}{k\electronvolt} \\ \SI{1000}{\micro\second/frame} \si{\times} \SI{60000}{frames} for \SI{5.415}{k\electronvolt} \\ (Initial 20 frames were excluded from analysis \\because of whole readout stability)} \\ \hline
 \shortstack{\strut Scan time \\ (Settling wait time \\for analog output)} & \SI{320}{ns/pixel} \\ \hline
 Sensor node reset & \SI{1}{\micro\second/frame} with \SI{300}{mV} ref. voltage \\ \hline
 CDS reset & \SI{2}{\micro\second/frame} with \SI{350}{mV} ref. voltage \\ \hline
 Readout board & SEABAS 2 \cite{seabas, ryunishiD, ryunishi3, ryunishi4} \\ \hline
 \end{tabular}
\caption{INTPIX4NA detector operation parameters for energy resolution measurements.}
\label{tab:exp_setup_detail_soi_2}
\end{table}

\begin{table}[H]
 \centering
 \begin{tabular}{|c|c|}
 \hline
 Identify Number & \shortstack{\strut Pixel Address \\(Column number, Row number)} \\ \hline
 1 & (18, 291) \\ \hline
 2 & (28, 291) \\ \hline
 3 & (38, 291) \\ \hline
 4 & (210, 291) \\ \hline
 5 & (220, 291) \\ \hline
 6 & (230, 291) \\ \hline
 7 & (402, 290) \\ \hline
 8 & (412, 290) \\ \hline
 9 & (422, 290) \\ \hline
 \end{tabular}
\caption{List of target pixels for energy resolution and gain linearity measurements.}
\label{tab:exp_targetpix}
\end{table}

Figure \ref{fig:energypeak_no8_5kev}, \ref{fig:energypeak_no8_8kev} and \ref{fig:energypeak_no8_12kev} are the distribution of summarized charge quantity of target pixel No.8, as the representative result. 
The results for all target pixels are shown in tables \ref{tab:exp_energypeak_5kev}, \ref{tab:exp_energypeak_8kev}, and \ref{tab:exp_energypeak_12kev}.
Energy resolution of target pixels are 35.3\%--46.2\% at \SI{5.415}{k\electronvolt}, 21.7\%--35.6\% at \SI{8}{k\electronvolt} and 15.7\%--19.4\% at \SI{12}{k\electronvolt}. 
The gains are 9.3--10.6 \si{\micro\volt/\elementarycharge}. 
In situations that involve using low-energy X-rays, such as in the case of \SI{5.415}{k\electronvolt}, the signal height of the X-ray and noises (thermal noise and readout noise) are very close under room temperature (approximately 25--27 \si{\degreeCelsius}) operation. 
Thus, the peak of one photon overlaps the background distribution. 
However, in the energy result for the case of \SI{5.415}{k\electronvolt}, we confirmed that X-ray signals can be separated from the noise under room temperature. 

\begin{table}[H]
 \centering
 \begin{tabular}{|c|c|c|}
 \hline
 \shortstack{\strut Identify \\Number} & \shortstack{\strut Energy \\resolution \\(\%)} & \shortstack{\strut Peak (ADU) \\(0.488mV/ADU)} \\ \hline
 1 & 35.3 & 27.8 \si{\pm{}} 4.2 (SD), FWHM 9.8 (1 photon) \\ \hline
 2 & 38.7 & 27.6 \si{\pm{}} 4.5 (SD), FWHM 10.7 (1 photon) \\ \hline
 3 & 46.2 & 26.9 \si{\pm{}} 5.3 (SD), FWHM 12.4 (1 photon) \\ \hline
 4 & 41.5 & 29.3 \si{\pm{}} 5.2 (SD), FWHM 12.2 (1 photon) \\ \hline
 5 & 38.1 & 30.1 \si{\pm{}} 4.9 (SD), FWHM 11.5 (1 photon) \\ \hline
 6 & 44.1 & 29.1 \si{\pm{}} 5.4 (SD), FWHM 12.8 (1 photon) \\ \hline
 7 & 35.5 & 30.8 \si{\pm{}} 4.6 (SD), FWHM 10.9 (1 photon) \\ \hline
 8 & 41.0 & 30.0 \si{\pm{}} 5.2 (SD), FWHM 12.3 (1 photon) \\ \hline
 9 & 42.2 & 31.1 \si{\pm{}} 5.6 (SD), FWHM 13.1 (1 photon) \\ \hline
 \end{tabular}
\caption{List of results of summarized charge quantity distribution at E = \SI{5.415}{k\electronvolt}.}
\label{tab:exp_energypeak_5kev}
\end{table}

\begin{table}[H]
 \centering
 \begin{tabular}{|c|c|c|}
 \hline
 \shortstack{\strut Identify \\Number} & \shortstack{\strut Energy \\resolution \\(\%)} & \shortstack{\strut Peak (ADU) \\(0.488mV/ADU)} \\ \hline
 1 & 28.0 & \shortstack{\strut 40.4 \si{\pm{}} 4.8 (SD), FWHM 11.3 (1 photon) \\ 82.4 \si{\pm{}} 5.4 (SD), FWHM 12.7 (2 photons) \\ 125.8 \si{\pm{}} 5.1 (SD), FWHM 12.0 (3 photons)} \\ \hline
 2 & 27.2 & \shortstack{\strut 42.8 \si{\pm{}} 4.9 (SD), FWHM 11.6 (1 photon) \\ 85.0 \si{\pm{}} 4.9 (SD), FWHM 11.5 (2 photons) \\ 127.6 \si{\pm{}} 5.1 (SD), FWHM 12.0 (3 photons)} \\ \hline
 3 & 26.1 & \shortstack{\strut 42.4 \si{\pm{}} 4.7 (SD), FWHM 11.1 (1 photon) \\ 85.0 \si{\pm{}} 6.3 (SD), FWHM 14.9 (2 photons) \\ 129.2 \si{\pm{}} 5.4 (SD), FWHM 12.8 (3 photons)} \\ \hline
 4 & 25.0 & \shortstack{\strut 44.6 \si{\pm{}} 4.7 (SD), FWHM 11.2 (1 photon) \\ 90.1 \si{\pm{}} 6.0 (SD), FWHM 14.2 (2 photons) \\ 133.5 \si{\pm{}} 4.5 (SD), FWHM 10.5 (3 photons)} \\ \hline
 5 & 29.0 & \shortstack{\strut 46.1 \si{\pm{}} 5.7 (SD), FWHM 13.3 (1 photon) \\ 91.7 \si{\pm{}} 6.6 (SD), FWHM 15.5 (2 photons) \\ 137.5 \si{\pm{}} 5.5 (SD), FWHM 12.9 (3 photons)} \\ \hline
 6 & 28.2 & \shortstack{\strut 45.0 \si{\pm{}} 5.4 (SD), FWHM 12.7 (1 photon) \\ 90.4 \si{\pm{}} 5.2 (SD), FWHM 12.2 (2 photons) \\ 135.0 \si{\pm{}} 5.6 (SD), FWHM 13.2 (3 photons)} \\ \hline
 7 & 35.6 & \shortstack{\strut 46.6 \si{\pm{}} 7.0 (SD), FWHM 16.6 (1 photon) \\ 92.8 \si{\pm{}} 3.9 (SD), FWHM 9.2 (2 photons) \\ 138.9 \si{\pm{}} 6.9 (SD), FWHM 16.1 (3 photons)} \\ \hline
 8 & 28.0 & \shortstack{\strut 46.0 \si{\pm{}} 5.5 (SD), FWHM 12.9 (1 photon) \\ 91.9 \si{\pm{}} 5.6 (SD), FWHM 13.2 (2 photons) \\ 139.5 \si{\pm{}} 5.7 (SD), FWHM 13.4 (3 photons)} \\ \hline
 9 & 21.7 & \shortstack{\strut 48.0 \si{\pm{}} 4.4 (SD), FWHM 10.4 (1 photon) \\ 94.8 \si{\pm{}} 4.4 (SD), FWHM 10.4 (2 photons) \\ 139.5 \si{\pm{}} 6.9 (SD), FWHM 16.1 (3 photons)} \\ \hline
 \end{tabular}
\caption{List of results of summarized charge quantity distribution at E = \SI{8}{k\electronvolt}.}
\label{tab:exp_energypeak_8kev}
\end{table}

\begin{table}[H]
 \centering
 \begin{tabular}{|c|c|c|}
 \hline
 \shortstack{\strut Identify \\Number} & \shortstack{\strut Energy \\resolution \\(\%)} & \shortstack{\strut Peak (ADU) \\(0.488mV/ADU)} \\ \hline
 1 & 19.4 & \shortstack{\strut 63.1 \si{\pm{}} 5.2 (SD), FWHM 12.3 (1 photon) \\ 127.7 \si{\pm{}} 3.8 (SD), FWHM 9.0 (2 photons) \\ 192.6 \si{\pm{}} 5.7 (SD), FWHM 13.4 (3 photons)} \\ \hline
 2 & 18.0 & \shortstack{\strut 65.5 \si{\pm{}} 5.0 (SD), FWHM 11.8 (1 photon) \\ 130.2 \si{\pm{}} 4.4 (SD), FWHM 10.4 (2 photons) \\ 195.8 \si{\pm{}} 2.9 (SD), FWHM 6.9 (3 photons)} \\ \hline
 3 & 17.5 & \shortstack{\strut 65.2 \si{\pm{}} 4.9 (SD), FWHM 11.4 (1 photon) \\ 130.5 \si{\pm{}} 5.4 (SD), FWHM 12.8 (2 photons) \\ 196.1 \si{\pm{}} 5.6 (SD), FWHM 13.3 (3 photons)} \\ \hline
 4 & 18.3 & \shortstack{\strut 68.1 \si{\pm{}} 5.3 (SD), FWHM 12.4 (1 photon) \\ 138.1 \si{\pm{}} 6.0 (SD), FWHM 14.1 (2 photons) \\ 206.4 \si{\pm{}} 6.2 (SD), FWHM 14.5 (3 photons)} \\ \hline
 5 & 17.3 & \shortstack{\strut 69.0 \si{\pm{}} 5.1 (SD), FWHM 11.9 (1 photon) \\ 138.4 \si{\pm{}} 6.2 (SD), FWHM 14.6 (2 photons) \\ 206.6 \si{\pm{}} 5.3 (SD), FWHM 12.4 (3 photons)} \\ \hline
 6 & 18.9 & \shortstack{\strut 69.9 \si{\pm{}} 5.6 (SD), FWHM 13.2 (1 photon) \\ 139.7 \si{\pm{}} 5.9 (SD), FWHM 13.9 (2 photons) \\ 208.5 \si{\pm{}} 4.4 (SD), FWHM 10.4 (3 photons)} \\ \hline
 7 & 16.4 & \shortstack{\strut 70.5 \si{\pm{}} 4.9 (SD), FWHM 11.5 (1 photon) \\ 141.9 \si{\pm{}} 4.3 (SD), FWHM 10.1 (2 photons) \\ 212.4 \si{\pm{}} 5.4 (SD), FWHM 12.7 (3 photons)} \\ \hline
 8 & 15.7 & \shortstack{\strut 71.5 \si{\pm{}} 4.8 (SD), FWHM 11.3 (1 photon) \\ 142.2 \si{\pm{}} 6.5 (SD), FWHM 15.3 (2 photons) \\ 214.3 \si{\pm{}} 6.2 (SD), FWHM 14.5 (3 photons)} \\ \hline
 9 & 16.2 & \shortstack{\strut 74.0 \si{\pm{}} 5.1 (SD), FWHM 12.0 (1 photon) \\ 145.8 \si{\pm{}} 6.1 (SD), FWHM 14.2 (2 photons) \\ 217.3 \si{\pm{}} 4.8 (SD), FWHM 11.2 (3 photons)} \\ \hline
 \end{tabular}
\caption{List of results of summarized charge quantity distribution at E = \SI{12}{k\electronvolt}.}
\label{tab:exp_energypeak_12kev}
\end{table}

\begin{figure}[H]
\centering
\includegraphics[width=\linewidth]{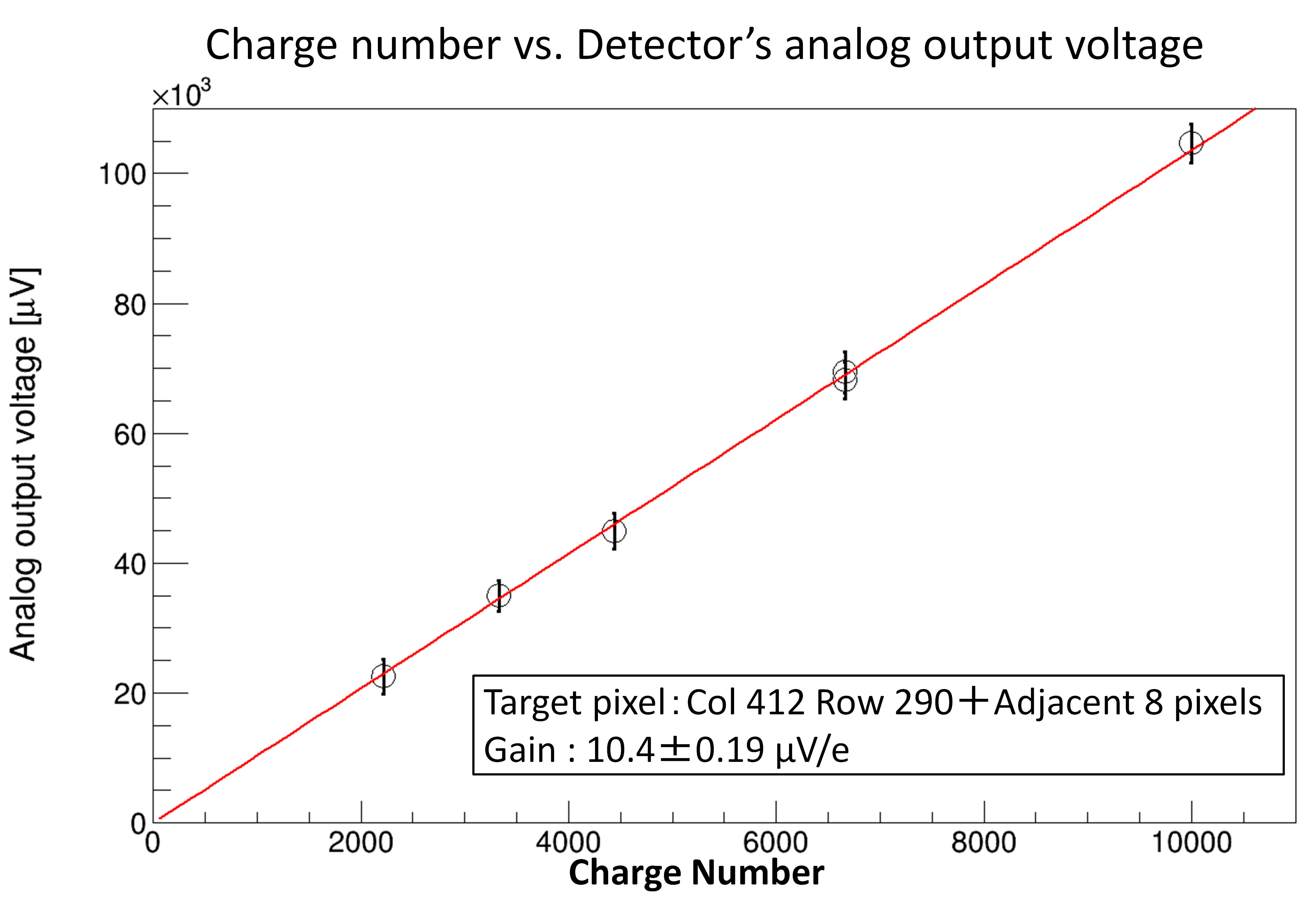}
\caption{Plot of the signal height at each peak and the estimated number of holes of target pixel No.8.}
\label{fig:exp_gain}
\end{figure}

In this study, the gain linearity was analyzed from the charge distribution at the 1--3 peaks of \SI{8} and \SI{12}{k\electronvolt} data from the energy resolution measurements shown above. 
Figure \ref{fig:exp_gain} is the result of target pixel No.8, the plot of the relationship between the signal height at each peak and the estimated number of charges, as the representative result. 
Results of all target pixels are shown in table \ref{tab:exp_gain}.
The gain value is slightly lower than the specific value, and there are some variations depending on the location. 
There is a possibility that each pixel node has some variation in capacitance depending on the structure of the pixel array and peripheral circuit. 
However, the results of each target show high linearity; thus, this can be corrected by a simple adjustment process. 

\begin{table}[H]
 \centering
 \begin{tabular}{|c|c|}
 \hline
 Identify Number & Gain (\si{\micro\volt/\elementarycharge}) \\ \hline
 1 & 9.3 \si{\pm{}} 0.16 \\ \hline
 2 & 9.5 \si{\pm{}} 0.12 \\ \hline
 3 & 9.5 \si{\pm{}} 0.18 \\ \hline
 4 & 10.0 \si{\pm{}} 0.18 \\ \hline
 5 & 10.1 \si{\pm{}} 0.18 \\ \hline
 6 & 10.1 \si{\pm{}} 0.16 \\ \hline
 7 & 10.3 \si{\pm{}} 0.17 \\ \hline
 8 & 10.4 \si{\pm{}} 0.19 \\ \hline
 9 & 10.6 \si{\pm{}} 0.17 \\ \hline
 \end{tabular}
\caption{List of the results of gain analysis.}
\label{tab:exp_gain}
\end{table}

\subsection{Minimum settling wait time}
\label{sec:exp3}
The INTPIX4NA uses a global shutter, and this shutter mode requires certain time to read out the signal from the pixel array, called the ``dead time". 
Then, the total measurement time is more than equal to the actual exposure time plus the dead time. 
The settling wait time of each pixel's analog output, most of the dead time, 
is defined as the required wait period from the timing of one pixel output connected to the series of output buffers to the timing of the sampling by the ADC. 
This is the main parameter limiting the frame rate; for example, if the scan time is extended by \SI{1}{ns}, one cycle of the readout procedure period will be extended by \SI{32768}{ns} (full array readout case). 
In INTPIX4NA, the settling time was improved from the original INTPIX4 to minimize the dead time by improving the circuit of the output buffers. 

The settling wait time (shown as ``Scan time" in the results) was tested at BL-14A using \SI{12}{k \\ electronvolt} monochromatic X-ray beams. 
The energy resolution was measured with a variable scan time from 80 to 640 ns/pixel. 
The procedure of the energy resolution measurement was the same as that described in Section \ref{sec:exp2}. 
The detector and optical setup were also almost the same as those of the energy resolution measurement shown above. 
Detector operation parameters are shown in table \ref{tab:exp_setup_detail_soi_3}. 

\begin{table}[H]
 \centering
 \begin{tabular}{|c|c|}
 \hline
 X-ray Energy & \SI{12}{k\electronvolt} monochromatic \\ \hline
 Sensor bias voltage & +\SI{250}{\volt} \\ \hline
 Sensor temperature & Room temperature (approximately 25--27 \si{\degreeCelsius}) \\ \hline
 Exposure time & \shortstack{\strut \SI{400}{\micro\second/frame} \si{\times} \SI{10000}{frames} \\ (Initial 20 frames were excluded from analysis \\because of whole readout stability)} \\ \hline
 \shortstack{\strut Scan time \\ (Settling wait time \\for analog output)} & \shortstack{\strut 80--640 \si{ns/pixel} \\ \SI{40}{ns/pixel} step \si{\times} 15 points} \\ \hline
 Sensor node reset & \SI{2}{\micro\second/frame} with \SI{300}{mV} ref. voltage \\ \hline
 CDS reset & \SI{2}{\micro\second/frame} with \SI{350}{mV} ref. voltage \\ \hline
 Readout board & SEABAS 2 \cite{seabas, ryunishiD, ryunishi3, ryunishi4} \\ \hline
 Target pixel & (412, 290) with adjacent 8 pixels \\ \hline
 \end{tabular}
\caption{INTPIX4NA detector operation parameters for settling time (Scan time) measurements.}
\label{tab:exp_setup_detail_soi_3}
\end{table}

\begin{figure}[H]
\centering
\includegraphics[width=\linewidth]{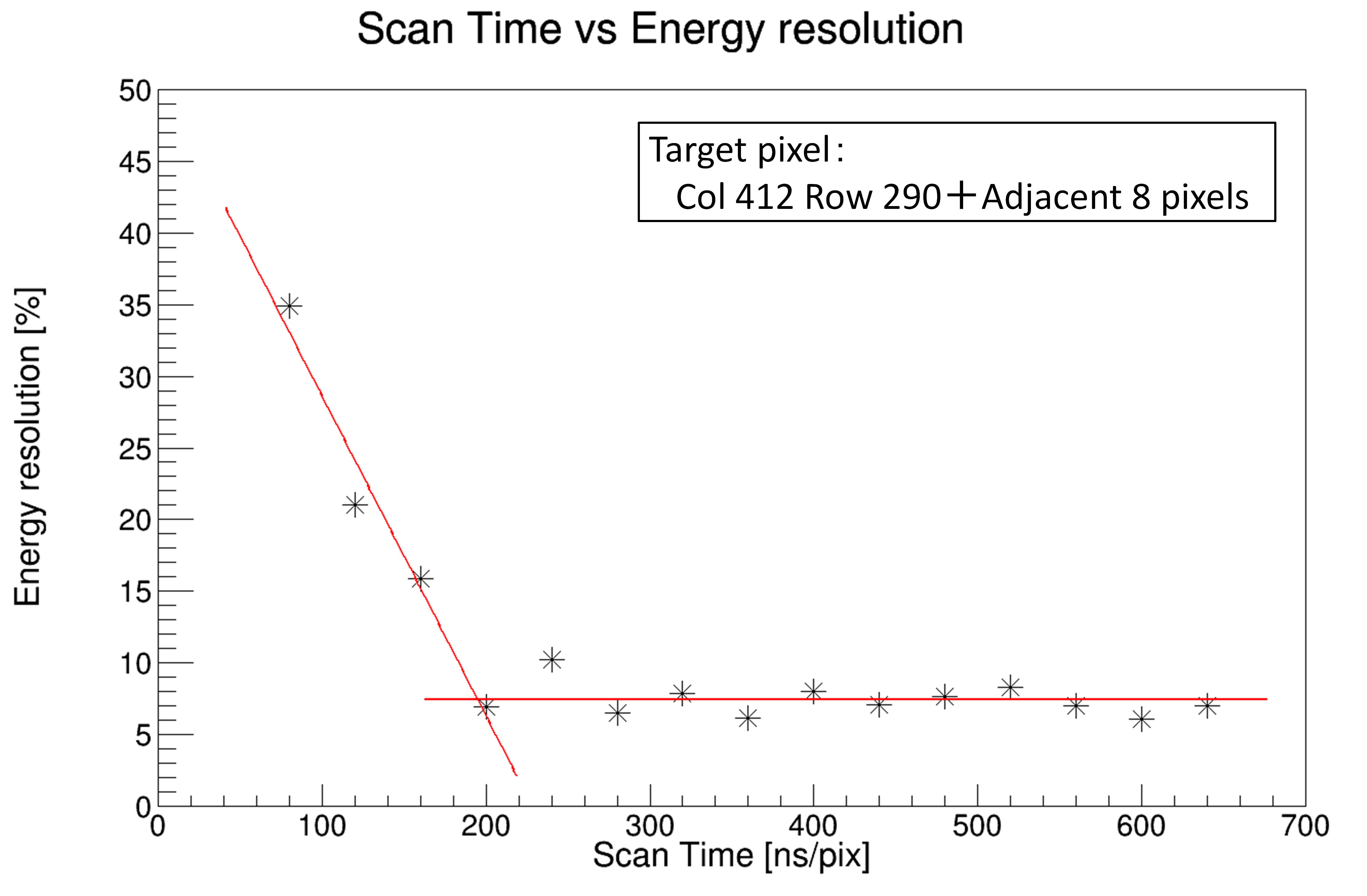}
\caption{Plot of settling time (Scan time) and Energy resolution of target pixel (412, 290) with adjacent 8 pixels.}
\label{fig:settling_no8}
\end{figure}

The plot of the settling time (scan time) measurement is shown in figure \ref{fig:settling_no8}. 
Energy resolution reached a plateau at scan time = 200 ns / pixel. 
This means that 153 fps (200 ns \si{\times} 64 \si{\times} 512, full-array with parallel readout mode) is the maximum frame rate of the current setup of this detector. 
However, if the degradation of energy resolution performance (35\% at \SI{12}{k\electronvolt}) is allowed, 80 ns/pixel operation is usable. 

This result includes the degradation caused by current SEABAS2 readout system. 
Thus, there is a possibility of reducing the settling wait time by improving the setup, such as by reducing the analog signal transfer length. 
Furthermore, the minimum settling wait time was mainly limited by the throughput of 1 gigabit Ethernet, longer than \SI{320}{ns/pixel} in full array mode. 
Thus, the settling wait time was measured with a 64 \si{\times} 64 pixel region of interest (ROI) mode and full array measurement will be planned with a newer readout system. 

\section{Conclusion}
\label{sec:conc}
We developed a new X-ray detector, INTPIX4NA, and performed several performance tests at the Photon Factory (KEK). 
These tests showed the following results. 
\begin{enumerate}
\setlength{\leftskip}{-10pt}
\setlength{\itemsep}{2pt}
\setlength{\parskip}{0pt}
\setlength{\itemindent}{0pt}
\setlength{\labelsep}{3pt}
\item INTPIX4NA's MTF shows more than 50\% at Nyquist frequency (\SI{29.4}{cycle/mm}) in both the vertical and horizontal direction. 
In addition, 50\% and 10\% MTF were beyond the Nyquist frequency in both the vertical and horizontal directions. 
These MTF characteristics reach the limitation of pixel size and have sufficient spatial resolution performance to measure with the setup of the camera length 15 mm or less. 
\item The energy resolution analyzed from the collected charge counts of target pixels are 35.3\%--46.2\% at \SI{5.415}{k\electronvolt}, 21.7\%--35.6\% at \SI{8}{k\electronvolt} and 15.7\%--19.4\% at \SI{12}{k\electronvolt}. 
The X-ray signal can be separated from the noise even at low-energy X-rays (\SI{5.415}{k\electronvolt}) at room temperature (approximately 25--27 \si{\degreeCelsius}). 
\item The gains are 9.3--10.6 \si{\micro\volt/\elementarycharge}. 
The gain value is slightly lower than the specific value, and there are some variations depending on the location. 
However, the results of each target show good linearity; thus, this can be corrected by a simple adjustment process. 
In total, the gain linearity is sufficient for the X-ray energies mainly used for residual stress measurements. 
\item The minimum settling time where the signal quality can be well maintained is 200 ns / pixel in the current measurement system. 
This is equivalent to a frame rate of 153 fps (full-array with parallel readout mode). 
If the degradation of energy resolution performance (35\% at \SI{12}{k\electronvolt}) is allowed, 80 ns/pixel operation (equivalent to 350 fps for a full array) is usable. 
\end{enumerate}
These results satisfy the required performance in the air and at room temperature (approximately 25--27 \si{\degreeCelsius}) condition that is assumed for the environment of the residual stress measurement. 
Thus, we consider that this detector is ready to use for the residual stress measurement. 
Moreover, currently, we are considering further studies mentioned below. 
\begin{enumerate}
\setlength{\leftskip}{-10pt}
\setlength{\itemsep}{2pt}
\setlength{\parskip}{0pt}
\setlength{\itemindent}{0pt}
\setlength{\labelsep}{3pt}
\item Integration into a new stress measurement system under development and evaluated its overall performance to meet the requirements of industrial standards \cite{cosastd}. 
\item High accuracy performance evaluation with a temperature controlled setup (less than \SI{-20}{\degreeCelsius} for low-energy X-rays). 
\item Evaluation of over 300 fps high speed imaging (settling time less than 100 ns / pixel with full array) with improved setup and faster readout system. 
\item Development of high accuracy pixel gain correction method which reduces the gain flatness variability to less than \si{\pm{}} 1\%. 
\item Enlargement of the sensitive area by tiling several chips (currently 4 chips tiling is planned) and its application to X-ray imaging measurements using lower energy such as X-ray micro computed tomography. 
\end{enumerate}

\acknowledgments

This study was conducted under the approval of the Photon Factory Program Advisory Committee (Proposal No. 2019G606, 2019G608 and 2020PF-33). 
This study was performed on behalf of SOIPIX collaboration \cite{soipix}. 
INTPIX4NA was developed under the collaboration of High Energy Accelerator Research Organization (KEK), Kanazawa University and NACHI-FUJIKOSHI CORP. 
The study was partly supported by the ISIJ Research Promotion Grant from the Iron and Steel Institute of Japan.


\end{document}